\documentclass[11pt]{article}
\usepackage{fullpage}
\usepackage{graphicx}
\usepackage{latexsym,amsmath,amssymb,amsthm,epic,eepic,multirow}
\usepackage{natbib}

\usepackage{multirow}
\usepackage{subfigure} 
\usepackage{natbib}
\usepackage{algorithm}
\usepackage{algorithmic}
\usepackage{mdframed}
\usepackage{tikz}
\usepackage{hyperref}
\usepackage{listings} 
\usepackage{color} 

\usepackage[toc,page]{appendix} 

\newenvironment{abstract2}
 {\small
  \begin{center}
  \bfseries Summary\vspace{-.5em}\vspace{0pt}
  \end{center}
  \quotation}
 {\endquotation}

\definecolor{upmaroon}{rgb}{0.48, 0.07, 0.07}
\definecolor{royalazure}{rgb}{0.0, 0.22, 0.66}
\definecolor{pakistangreen}{rgb}{0.0, 0.4, 0.0}

\newtheorem{Corollary}{Corollary}

\newtheorem{Lemma}{Lemma}

\newtheorem{Theorem}{Theorem}
 
\newtheorem{Remark}{Remark}
\newtheorem{Assumption}{Assumption}

\newcommand{\argmin}{\arg\min}

\newcommand{\RR}{\mathbf{R}}
\newcommand{\R}{\mathbb{R}}
\newcommand{\E}{\mathbf{E}}

\newcommand{\A}{A}
\newcommand{\QW}{{\rm Q}_{\Sigma}}
\newcommand{\QWh}{\widehat{{\rm Q}}_{\Sigma}} 
\newcommand{\QA}{{\rm Q}_{\A} }
\newcommand{\QAh}{\widehat{{\rm Q}}_{\A} } 
\newcommand{\VW}{{\rm V}_{\Sigma}}
\newcommand{\VWh}{\widehat{{\rm V}}_{\Sigma}(\tau)} 
\newcommand{\VA}{{\rm V}_{\A}}
\newcommand{\VAh}{\widehat{{\rm V}}_{\A}(\tau)} 
\newcommand{\V}{{\rm V}}
\newcommand{\BW}{{\rm B}_{\Sigma} }
\newcommand{\BA}{{\rm B}_{\A} }
\newcommand{\MW}{{\rm M}_{\Sigma} }
\newcommand{\MA}{{\rm M}_{\A} }

\newcommand{\set}{G}

\newcommand{\cip}{\overset{p}{\to}}
\newcommand{\cid}{\overset{d}{\to}}
\newcommand{\scale}{(1+\eta)}

\definecolor{Igreen}{rgb}{0.0, 0.56, 0.0}

\begin{document}

\title{Group Inference in High Dimensions with Applications to
  Hierarchical Testing} 
\author{Zijian Guo, \; Claude Renaux, \; Peter B\"{u}hlmann  \; and\;  T. Tony Cai} 

\date{\today}

\maketitle

\begin{abstract2}
High-dimensional group inference is an essential part of statistical methods for analysing complex data sets, including hierarchical testing, tests of interaction, detection of heterogeneous treatment effects and inference for local heritability.  Group inference in regression models can be measured with respect to a weighted quadratic functional of the regression sub-vector corresponding to the group. Asymptotically unbiased estimators of these
weighted quadratic functionals are constructed and a novel procedure using these estimators for inference  is proposed.  We derive its asymptotic Gaussian distribution which enables the construction of asymptotically valid confidence intervals and tests which perform well in terms of length or power. The proposed test is computationally efficient even for a large group, statistically valid for any group size and achieving good power performance for testing large groups with many small regression coefficients. We apply the methodology to several interesting statistical problems and
demonstrate its strength and usefulness on simulated and real data.\\

\noindent\small{{\bf Key words:} Heterogeneous Effects; Interaction Test;  Local Heritability; Debiasing Lasso; Partial regression.}
\end{abstract2}

\section{Introduction}

\subsection{Motivation and Formulation}
Statistical inference for high-dimensional linear regression is an important but also challenging problem. 
This paper addresses a long-standing statistical problem, {namely inference or testing
  significance of groups of covariates.}
Specifically, we consider the following high-dimensional linear regression 
\begin{equation}
y_i=X_{i\cdot}^{\intercal}\beta+\epsilon_i, \quad \text{for} \; i = 1, \ldots, n,
\label{eq: linear model}
\end{equation}
{where $X_{i\cdot}\in \RR^{p}$ are independent and identically distributed random vectors with $\Sigma=\E X_{i\cdot}X_{i\cdot}^{\intercal}$  and $\epsilon_i$ are  independent and identically distributed centred Gaussian errors, independent of $X_{i\cdot}$, with variance $\sigma^2$.} 
For a given index set $\set \subseteq
\left\{1,2,\ldots,p\right\},$ we consider testing $H_0: \beta_{G}=0$ through testing the equivalent null hypothesis,
\begin{equation}
{H}_{0,A}: \; \beta_{\set}^{\intercal}A\beta_{\set}=0, 
\label{eq: group testing}
\end{equation}
where $A\in \R^{|G|\times |G|}$ is a positive-definite matrix with $|G|$ denoting the cardinality of $G$. 
 The test \eqref{eq: group testing} includes a class of tests for different $A$.
By taking $A$ as $\Sigma_{G,G}$, \eqref{eq: group testing} is specifically written as 
\begin{equation}
{H}_{0,\Sigma}: \; \beta_{\set}^{\intercal}\Sigma_{\set,\set}\beta_{\set}=0.
\label{eq: group testing a}
\end{equation}
In addition, when $A$ is the identity matrix, \eqref{eq: group testing} is reduced to ${H}_{0,{\rm I}}: \; \beta_{\set}^{\intercal}\beta_{\set}=0.$
The quantity $\beta_{\set}^{\intercal}\Sigma_{\set,\set}\beta_{\set}$ in
\eqref{eq: group testing a} is naturally used for group significance
testing as the quantity itself measures the variance explained by the set
of variables $X_{i,G},$ 
$
\beta_{\set}^{\intercal}\Sigma_{\set,\set}\beta_{\set}=\E |X_{i,G}^{\intercal}\beta_{G}|^2.
$
 {The testing problem \eqref{eq: group testing a} can be conducted in both settings where the matrix $\Sigma_{G,G}$ is known or unknown. If $\Sigma_{G,G}$ is known, then it can be simply treated as a special case of  \eqref{eq: group testing}. 
However, in the more practical setting with  an unknown $\Sigma_{G,G}$, we need to estimate $\Sigma_{G,G}$ in the construction of the test statistic and quantify the additional uncertainty of estimating this matrix from data.} 

In the following, we shall provide a series of motivations for {group inference or 
  significance.} 

\noindent{\bf 1. Hierarchical Testing.} It
is often too ambitious to detect significant single variables and groups
of correlated variables are considered instead. 
Hierarchical testing is adjusting hierarchically over multiple group tests. It
 addresses the trade-off between
signal strength and high correlation and it is a most
natural way to deal with large-scale high-dimensional testing problems
in real applications \citep{buzduganetal16, klasen2016multi}. More
details are given in Section \ref{sec: hier appl}.\\
 \vspace{-4mm}

\noindent{\bf 2. Interaction Test and Detection of Effect Heterogeneity.} Group
  significance testing can be used to examine the existence of interaction. We write the model
  with interaction terms \citep{tian2014simple,cai2019individualized}, 
$y_i=X_{i\cdot}^{\intercal}\beta+D_{i}\cdot X_{i\cdot}^{\intercal}\gamma+\epsilon_i$ for $i=1, \ldots, n$
with $D_i$ denoting the exposure variable and formulate
the interaction test as $H_0: \gamma=0$.  If $D_i$ denotes whether the
$i$-th observation receives a treatment, then one can test against the presence of heterogenous treatment effects, see also Section \ref{sec: interaction}.\\

 \vspace{-4mm}

\noindent{\bf 3. Local Heritability {in Genetics}.}  
  Local heritability is among the most
  important 
  heritability measures \citep{shi2016contrasting} and can be defined as $\E |X_{i,G}^{\intercal}\beta_{G}|^2=\beta_{\set}^{\intercal}\Sigma_{\set,\set}\beta_{\set}$ in \eqref{eq: linear model}, representing the
  proportion of variance explained by a subset of genotypes indexed by the
  group $G$. In applications, the group $G$ can be naturally formulated,
  e.g., the set of SNPs located on the
  same chromosome. It is of interest to test whether a group of genotypes with the index set $G$ is significant and construct confidence intervals for local heritability.

\subsection{Results and Contribution}
Statistical inference in high dimensional models has been studied in both statistics and econometrics with a focus on confidence interval  and hypothesis testing
for individual regression coefficients
\citep{javanmard2014confidence,van2014asymptotically,zhang2014confidence,chernozhukov2017central,cai2015regci}. Together with a careful use of
bootstrap methods, certain maximum tests have been developed in
\cite{chernozhukov2017central,dezeure2017high,zhang2017simultaneous} to
conduct the hypothesis testing problem $H_0: \beta_{G}=0$. These tests rely on the maximum of individual estimates and are tailored for sparse alternatives, and they can be computationally costly, especially when the size of $G$ is large.  
Instead of the maximum statistics which is favourable for sparse alternatives, we focus on sum-type statistics and also address the computational issue through directly testing the group significance.
In low dimensions, {the (partial)} F-test {is the classical sum-type procedure} for testing group significance. However, there is {a} lack of sum-type methods for conducting group significance tests in high dimensions, especially for large groups. Our proposed group test with sum-type statistics can hence be viewed as an additional potentially powerful procedure in the toolkit of high-dimensional data analysis. 

The proposed test statistic is constructed via an inference procedure for the quadratic
form $\beta_{\set}^{\intercal}A\beta_{\set}$ in \eqref{eq: group testing a}.  We illustrate the main idea using 
$\QW=\beta_{\set}^{\intercal}\Sigma_{G,G}\beta_{\set}$ as an example.
Denote by $\widehat{\beta}$ a reasonably good estimator (e.g. the Lasso estimator) of $\beta$ and define $\widehat{\Sigma}=\sum_{i=1}^{n}X_{i\cdot}X_{i\cdot}^{\intercal}/n$.
The plug-in estimator $\widehat{\beta}^{\intercal}_{G}\widehat{\Sigma}_{G,G}\widehat{\beta}_{G}$
is not proper for statistical inference because it has a dominating bias inherited from $\widehat{\beta}.$
We correct the bias of this plug-in estimator through a novel projection.
The proposed methodology can be extended to deal with the more general inference problem for $\beta_{\set}^{\intercal}A\beta_{\set}$ 
where $A\in \R^{|G|\times |G|}$ is a positive definite matrix. 

The proposed test is valid for any group size $|G|$ in terms of type-I error control.  As a generalization of the F-test to high-dimensions, 
the group tests proposed in \cite{mitra2016benefit} and \cite{van2016chi} require the group size $|G|$ to be smaller than the sample size. 
Our test has good power performance for testing large groups. The proposed test  is asymptotically powerful as long as $\beta_{\set}^{\intercal}\Sigma_{G,G}\beta_{\set}$ is of a larger order of magnitude than $(1+\|\beta_{G}\|_{2}+\|\beta_{G}\|_2^2)/n^{1/2}$. In comparison, the detection threshold of the $\chi^2$ test is in the order of $(|G|/n)^{1/2}$ \citep{mitra2016benefit,van2016chi} which is inferior to the proposed test for a large $|G|.$
Additionally, for certain difficult settings, our proposed test achieves the same detection boundaries as the maximum test \citep{chernozhukov2017central,dezeure2017high,zhang2017simultaneous}, up to a constant, see the discussion after Corollary \ref{cor: 2}.

In Section \ref{sec: simulation}, we compare the finite sample performance of the proposed sum-type test with the maximum test. In simulated data where the regression vector has many non-zero and small entries, we have observed that the proposed test has a better power performance; in simulated data with high correlation among covariates, the proposed confidence intervals achieve the desired coverage property while the coordinate-based inference procedures exhibit under-coverage. 

Based on the proposed group test, a hierarchical testing algorithm inherits all of the above advantages in terms of both the computational efficiency and statistical validity. Even conceptually, hierarchical testing is fundamentally requiring a group test that is not based on the maximum of individual coordinates: the latter simply corresponds to a Bonferroni adjustment of individual coordinate tests and a hierarchical testing procedure would be useless. We illustrate some practical results in Sections \ref{sec: sim hier} and \ref{sec: colony}.

\subsection{Literature Comparison}
In addition to the existing work based on the maximum test and generalization of $F$ tests mentioned above, there is other related work. Inference for the quadratic functionals is closely related to the group test. Statistical inference for
$\beta^{\intercal}\Sigma\beta$ and $\|\beta\|_2^2$ has been carefully
investigated in \cite{verzelen2018adaptive,cai2018semi,guo2019optimal} and the methods
developed in \cite{cai2018semi} have been extended to signal detection, which is a special case of  \eqref{eq: linear model} by
setting $G=\{1,\ldots,p\}$. The group
significance test problem is more challenging than signal detection, mainly due to the fact that the group test
requires a decoupling between variables in $G$ and $G^{c}$. The decoupling between $G$ and $G^{c}$ essentially
requires a novel construction of the projection direction. See Remark \ref{rem: comparison-quad} in Section \ref{sec: method} for more details. The same
comments can be made to differentiate the current paper with the signal
detection problem considered in
\cite{ingster2010detection,arias2011global}. Finally, \cite{meinshausen2015group} proposes another group significance test without
any compatibility condition on the design. The price to be paid for this relaxation of conditions is in terms of a typically large drop in power.  

\vspace{-10pt}
\section{Methodology for Testing Group Significance}
\label{sec: method}
We first present an inference procedure for $\QW=\beta_{\set}^{\intercal}\Sigma_{\set,\set}\beta_{\set}$ and will generalize it to inference for $\QA=\beta_{\set}^{\intercal}A{\beta_{\set}}.$ Throughout the paper, we use $\widehat{\beta}$ to denote a reasonably good estimator of $\beta$ and use $\widehat{\Sigma}=\sum_{i=1}^{n}X_{i\cdot}X_{i\cdot}^{\intercal}/n$ as the estimator of $\Sigma$. Without loss of generality, we assume the index set $G$ of the form $\{1,\ldots,|G|\}.$  To estimate $\QW$, we examine the error decomposition of a plug-in estimator $\widehat{\beta}_{\set}^{\intercal}\widehat{\Sigma}_{\set,\set}\widehat{\beta}_{\set}$ as 
$\widehat{\beta}_{\set}^{\intercal}\widehat{\Sigma}_{\set,\set}\widehat{\beta}_{\set}-\beta_{\set}^{\intercal}\Sigma_{\set,\set}{\beta_{\set}}=-2\widehat{\beta}_{\set}^{\intercal}\widehat{\Sigma}_{\set,\set}(\beta_{\set}-\widehat{\beta}_{\set})+\beta_{\set}^{\intercal}(\widehat{\Sigma}_{\set,\set}-\Sigma_{\set,\set})\beta_{\set}
-(\widehat{\beta}_{\set}-\beta_{\set})^{\intercal}\widehat{\Sigma}_{\set,\set}(\widehat{\beta}_{\set}-\beta_{\set}).$ In this decomposition, we need to estimate $2\widehat{\beta}_{\set}^{\intercal}\widehat{\Sigma}_{\set,\set}(\beta_{\set}-\widehat{\beta}_{\set})$ {as this is the dominant term} to further calibrate the plug-in estimator $\widehat{\beta}_{\set}^{\intercal}\widehat{\Sigma}_{\set,\set}\widehat{\beta}_{\set}.$ The calibration can be done through identifying a projection direction $\widehat{u}\in \R^{p}$ to approximate $\widehat{\beta}_{\set}^{\intercal}\widehat{\Sigma}_{\set,\set}(\beta_{\set}-\widehat{\beta}_{\set}).$ For any $u\in \R^{p},$
$$
u^{\intercal}\frac{1}{n}X^{\intercal}(y-X\widehat{\beta})-\widehat{\beta}_{\set}^{\intercal}\widehat{\Sigma}_{\set,\set}(\beta_{\set}-\widehat{\beta}_{\set})=\frac{1}{n} u^{\intercal}X^{\intercal}\epsilon+\left\{\widehat{\Sigma}u -(\widehat{\beta}_{\set}^{\intercal}\widehat{\Sigma}_{\set,\set}, {\bf 0})^{\intercal}\right\}^{\intercal}(\beta-\widehat{\beta}).
$$
In this decomposition, $\frac{1}{n} u^{\intercal}X^{\intercal}\epsilon$ can be viewed as a variance component with asymptotically normal distribution; the second term on the right hand side can be controlled by H{\"o}lder inequality, $\Big|\Big\{\widehat{\Sigma}u -\big(\widehat{\beta}_{\set}^{\intercal}\widehat{\Sigma}_{\set,\set}, {\bf 0}\big)^{\intercal}\Big\}^{\intercal}(\beta-\widehat{\beta})\Big|\leq \|\beta-\widehat{\beta}\|_1\Big\|\widehat{\Sigma}u -(\widehat{\beta}_{\set}^{\intercal}\widehat{\Sigma}_{\set,\set}, {\bf 0})^{\intercal}\Big\|_{\infty}.$ As long as $\widehat{\beta}$ is a reasonable estimator with a small $\|\beta-\widehat{\beta}\|_1,$ it remains to construct the projection direction $u\in \R^{p}$ such that $\big\|\widehat{\Sigma}u -(\widehat{\beta}_{\set}^{\intercal}\widehat{\Sigma}_{\set,\set}, {\bf 0})^{\intercal}\big\|_{\infty}$ is constrained.  A direct generalization of ``constraining bias and minimizing variance" in 
\citet{javanmard2014confidence} leads to
\begin{equation}
\begin{aligned}
\widetilde{u}=\arg\min \; u^{\intercal}\widehat{\Sigma} u  \quad
{\rm s.t.} \;\;  \| \widehat{\Sigma}u -(\widehat{\beta}_{\set}^{\intercal}\widehat{\Sigma}_{\set,\set}, {\bf 0})^{\intercal}\|_\infty \leq \|\widehat{\Sigma}_{\set,\set}\widehat{\beta}_{\set}\|_2 \lambda_{n}\\
\end{aligned}
\label{eq: projection ext}
\end{equation}
where $\lambda_{n}=C_0(\log p/n)^{1/2}$ for some constant $C_0>0.$  However, such a generalization does not always work: if $\|\widehat{\Sigma}_{\set,\set}\widehat{\beta}_{\set}\|_2 \lambda_{n}\geq \|\widehat{\Sigma}_{\set,\set}\widehat{\beta}_{\set}\|_{\infty}$, we have $\widetilde{u}=0$ as the solution and do not conduct any bias correction. When $|G|$ is large and the elements of the vector $\widehat{\Sigma}_{\set,\set}\widehat{\beta}_{\set}$ are of a similar order, then $\widetilde{u}=0$ as long as $\sqrt{|G|\log p/n}\geq C$ for some positive constant $C>0.$ 

To do the bias correction for arbitrary groups, we construct the projection direction $u$ as
\begin{equation}
\begin{aligned}
\widehat{u}=\arg\min \; u^{\intercal}\widehat{\Sigma} u  \quad
{\rm s.t.} \;\; & \max_{w\in \mathcal{C}}\left\langle w, \widehat{\Sigma}u -(\widehat{\beta}_{\set}^{\intercal}\widehat{\Sigma}_{\set,\set}, {\bf 0})^{\intercal}\right\rangle \leq \|\widehat{\Sigma}_{\set,\set}\widehat{\beta}_{\set}\|_2 \lambda_{n}
\end{aligned}
\label{eq: projection Cov}
\end{equation}
where $\lambda_{n}=C_0(\log p/n)^{1/2}$ for some constant $C_0>0$, and 
\begin{equation}
\mathcal{C}=\{e_1,\ldots,e_p,(\widehat{\beta}_{\set}^{\intercal}\widehat{\Sigma}_{\set,\set}/\|\widehat{\Sigma}_{\set,\set}\widehat{\beta}_{\set}\|_2, {\bf 0})^{\intercal}\}.
\label{eq: constraint set}
\end{equation}
In comparison to \eqref{eq: projection ext}, an additional direction $(\widehat{\beta}_{\set}^{\intercal}\widehat{\Sigma}_{\set,\set}, {\bf 0})^{\intercal}$ in $\mathcal{C}$ is introduced to ensure that the projection works for any group size and the main intuition of this additional constraint is to ensure the variance component dominates the remaining bias after the bias correction. Particularly, it rules out the trivial solution for a large $|G|.$
The final estimator of $\beta_{\set}^{\intercal}\Sigma_{\set,\set}\beta_{\set}$ is 
\begin{equation}
\QWh=\widehat{\beta}_{\set}^{\intercal}\widehat{\Sigma}_{\set,\set}\widehat{\beta}_{\set}+\frac{2}{n}\widehat{u}^{\intercal}X^{\intercal}(y-X\widehat{\beta}) \quad \text{with}\quad \widehat{\Sigma}=\frac{1}{n} X^{\intercal} X.
\label{eq: estimator Cov}
\end{equation}
We estimate $\sigma^2$ by $\widehat{\sigma}^2=\|y-X\widehat{\beta}\|_2^2/n$ and estimate the variance of $\QWh$ by
\begin{equation}
\VWh=\frac{4\widehat{\sigma}^2}{n}\widehat{u}^{\intercal}\widehat{\Sigma}\widehat{u}+ {\frac{1}{n^2}\sum_{ i=1}^{n} \left(\widehat{\beta}_{G}^{\intercal} X_{iG} X_{iG}^{\intercal}\widehat{\beta}_{G}-\widehat{\beta}_{G}^{\intercal}\widehat{\Sigma}_{G,G}\widehat{\beta}_{G}\right)^2}+\frac{\tau}{n},
\label{eq: variance Cov}
\end{equation}
for some positive constant $\tau>0$. We propose the $\alpha$-level test for $H_{0,\Sigma}$: \begin{equation}
\phi_{\Sigma}(\tau)=\mathbf{1}\left(\QWh\geq \scale {z_{1-\alpha}} (\VWh)^{1/2}
\right), 
\label{eq: test Cov}
\end{equation}
where $z_{1-\alpha}$ is the $1-\alpha$ quantile of the standard normal distribution and $\eta>0$ is a small constant and is typically set to $0.1$. A $(1-\alpha)$-level confidence interval for $\QW$ is given by
\begin{equation}
 {\rm CI}_{\Sigma}(\tau)=\left(\QWh-\scale{z_{1-\frac{\alpha}{2}}} (\VWh)^{1/2}, \,\, \QWh+\scale{z_{1-\frac{\alpha}{2}}} (\VWh)^{1/2}\right)\\ 
\label{eq: CI Cov}
\end{equation}
{We briefly discuss now the inclusion of $\tau$ and $\eta$ in \eqref{eq: test Cov}. The constant $\tau$ is essential to deal with
super-efficiency issues, see also the discussion after Theorem \ref{thm: Sigma middle}. The value $\eta$ is
solely used for more reliable finite sample performance controlling the type I error for situations where the
level of sparsity $k=\|\beta\|_0$ might be too large while being valid asymptotically as the assumption $k \leq c n^{1/2}/{\log p}$ in Theorem \ref{thm: Sigma middle}.}

\begin{Remark}
The computational cost of the estimator in \eqref{eq: estimator Cov} does not depend on the group size $|G|$. The construction of the projection direction $\widehat{u}$ involves solving a constrained optimization problem, or its dual penalized optimization problem, with a $p$-dimensional parameter. We de-bias $\widehat{\beta}_{\set}^{\intercal}\widehat{\Sigma}_{\set,\set}\widehat{\beta}_{\set}$ directly, instead of coordinate-wise de-biasing $\widehat \beta$. In comparison, the maximum test based on the debiased Lasso \citep{van2014asymptotically} or its bootstrap version \citep{chernozhukov2017central,dezeure2017high,zhang2017simultaneous} requires solving $|G|+1$ optimization problems. When $|G|$ is large, the computational improvement can be significant. See Table \ref{tab: time}.
\label{rem: computation}
\end{Remark}

\begin{Remark} 
The group significance test for arbitrary group size is more challenging than inference for $\|\beta\|_2^2$ in \cite{guo2019optimal} and inference for $\beta^{\intercal}\Sigma\beta$ in  \cite{cai2018semi} or \cite{verzelen2018adaptive}. 
Specifically, the additional constraint in \eqref{eq: constraint set} is not needed at all for estimating $\|\beta\|_2^2$ \citep{guo2019optimal}.
For $\beta^{\intercal}\Sigma\beta$, $\widehat{u}$ can be set as $\widehat{\beta}$ without even solving an optimization problem \citep{cai2018semi}. 
\label{rem: comparison-quad}
\end{Remark}

\begin{Remark} 
The proposed sum-type test is more immune to high correlation among the covariates than the maximum coordinate based test. It is worth noting that both the sum-type test and maximum coordinate test need a reasonably good initial estimator and in theory, this requires that the high-dimensional covariates are not highly correlated; See Assumptions \ref{assumption1} and \ref{assumption2}. Hence, robustness to high correlation of the sum-type test only happens at the bias-correction part, instead of the whole procedure.  In the special setting where  $X_{i,G}$ and $X_{i,G^{c}}$ are independent, we can construct 
$\widehat{u}=(\widehat{\beta}_{\set}^{\intercal}, {\bf 0})^{\intercal}$ and hence de-biasing $\widehat{\beta}_{\set}^{\intercal}\widehat{\Sigma}_{\set,\set}\widehat{\beta}_{\set}$ does not require the inversion of $\widehat{\Sigma}_{GG}$, which is useful when the covariates in $X_{i,G}$ are highly correlated. 
 In comparison, bias correction for constructing debiased estimators of $\beta$, on which the (bootstrapped) maximum test is based, tends to suffer from the high correlations. This observation shows that the proposed test is more reliable in high-correlation settings.  Additionally, $\QW=\E |X_{i,G}^{\intercal}\beta_{G}|^2$ accounts for the variance of the regression surface with $\tilde{y}_i=y_{i}-X_{i,G^{c}}^{\intercal}\beta_{G^{c}}.$ Therefore, $\QW$ is identifiable even if the components of $X_{i,G}$ exhibit high correlations. See Section \ref{sec: sim high-cor} for the numerical comparison.
\end{Remark}

The main idea of estimating $\QA=\beta_{\set}^{\intercal}A{\beta_{\set}}$ is similar to that of estimating $\QW$ though the problem itself is slightly easier due to the fact that the matrix $A$ is known. 
We start with the error decomposition of the plug-in estimator $\widehat{\beta}_{\set}^{\intercal}A\widehat{\beta}_{\set}$,
$
\widehat{\beta}_{\set}^{\intercal}A\widehat{\beta}_{\set}-\beta_{\set}^{\intercal}A{\beta_{\set}}=-2\widehat{\beta}_{\set}^{\intercal}A(\beta_{\set}-\widehat{\beta}_{\set})
-(\widehat{\beta}_{\set}-\beta_{\set})^{\intercal}A(\widehat{\beta}_{\set}-\beta_{\set}).
$
Similarly, we can construct the projection direction $\widehat{u}_{A}$ as 
$$
\widehat{u}_{A}=\arg\min \;\; u^{\intercal}\widehat{\Sigma} u \;\; {\rm s.t.} \;\;  \max_{w\in \mathcal{C}_{A}}\left\langle w, \widehat{\Sigma}u -\begin{pmatrix} \widehat{\beta}_{\set}^{\intercal}A& {\bf 0}\end{pmatrix}^{\intercal}\right\rangle \leq \|A\widehat{\beta}_{\set}\|_2 \lambda_{n}
$$
where $\lambda_{n}=C \sqrt{\log p/n}$ and 
$
\mathcal{C}_{A}=\left\{e_1,\ldots,e_p,(\widehat{\beta}_{\set}^{\intercal}A/ \|A\widehat{\beta}_{\set}\|_2, {\bf 0})^{\intercal}\right\}.$
Then we propose the final estimator of $\QA$ as 
\begin{equation}
\QAh=\widehat{\beta}_{\set}^{\intercal}A\widehat{\beta}_{\set}+2\widehat{u}_{A}^{\intercal}X^{\intercal}(y-X\widehat{\beta})/n
\label{eq: general A estimator}
\end{equation}
and  estimate the variance of $\QWh$ by $\VAh$ with
$
\VAh={4\widehat{\sigma}^2}\widehat{u}_{A}^{\intercal}\widehat{\Sigma}\widehat{u}_{A}/n+{\tau}/{n}
$
for some positive constant $\tau>0$. 
Having introduced the point estimators and the quantification of the variance, we propose the following two $\alpha$-level significance tests using a small $\eta>0$ (typically $\eta=0.1$):
{\begin{equation}
\phi_{\A}(\tau)=\mathbf{1}\left(\QAh\geq \scale{z_{1-\alpha}} (\VAh)^{1/2} \right),
\label{eq: test general}
\end{equation}
}
and construct CI as 
\begin{equation}
{\rm CI}_{\A}(\tau)=\left(\QAh-\scale{z_{1-\frac{\alpha}{2}}}(\VAh)^{1/2}, \,\, \QAh+\scale{z_{1-\frac{\alpha}{2}}} (\VAh)^{1/2}\right).
\label{eq: CI general}
\end{equation}

\vspace{-10pt}
\section{Theoretical Justification}
\label{sec: theory}
We first introduce the following regularity conditions before stating the main results.
\begin{Assumption}
\label{assumption1}
The rows $X_{i,\cdot} \in \RR^p$ are independent and identically distributed {sub-Gaussian} random vectors with $\Sigma=\E (X_{i,\cdot}X_{i,\cdot}^{\intercal} $) satisfying  $c_0\leq \lambda_{\min}\left(\Sigma\right) \leq \lambda_{\max}\left(\Sigma\right) \leq C_0$ for constants $C_0>c_0>0$. The errors $\epsilon_{1}, ..., \epsilon_{n}$ are independent and identically distributed centred Gaussian variables with variance $\sigma^2$ and are independent of $X$. 
\end{Assumption}

\begin{Assumption}
\label{assumption2}
With probability larger than $1-p^{-c}-\exp(-cn)$ for some positive constant $c>0$, the initial estimator $\widehat{\beta}$ and $\widehat{\sigma}^2$ satisfy,
 {\small
 \begin{equation*}
\|\widehat{\beta}-\beta\|_2 \leq  C ({\|\beta\|_0\log p}/{n})^{1/2}, \; \|\widehat{\beta}-\beta\|_1\leq C\|\beta\|_0({{\log p}/{n}})^{1/2}, \; \left|{\widehat{\sigma}^2}/{\sigma^2}-1\right| \leq C({1}/{{n}^{1/2}}+{\|\beta\|_0\log p}/{n}).
	\end{equation*}
	}for some positive constant $C>0.$
\end{Assumption}
\vspace{-2.5mm}
\begin{Assumption}
\label{assumption3}
The initial estimator $\widehat{\beta}$ is independent of $(X,y)$ used in the construction of \eqref{eq: estimator Cov} and \eqref{eq: general A estimator}. (For example by using sample splitting, see below).
\end{Assumption}
Assumption \ref{assumption1} implies the restricted eigenvalue condition introduced in \cite{bickel2009simultaneous} under the sparsity condition $\|\beta\|_0\leq c n/\log p$ for some positive constant $c>0$;
see \cite{zhou2009restricted,raskutti2010restricted} for the exact statement. The Gaussianity of $\epsilon_{i}$ is imposed to simplify the analysis and it can be weakened to sub-Gaussianity using a more refined analysis. 

Most of the high-dimensional estimators proposed in the literature satisfy the above Assumption \ref{assumption2} under regularity and sparsity conditions. See \cite{sun2012scaled,belloni2011square,bickel2009simultaneous} and the references therein for more details. 

Assumption \ref{assumption3} is imposed for technical analysis. 
With such an independence assumption, the asymptotic normality of the proposed estimators is easier to establish. However, such a condition is believed to be only technical and not necessary for the proposed method. As shown in the simulation study, we demonstrate that the proposed method, even not satisfying the independence assumption imposed in Assumption \ref{assumption3}, still works well numerically. 
We can also use  sample splitting to create the independence. We randomly split the data into two subsamples  $(X^{(1)},y^{(1)})$ with sample size $n_1=\lfloor {n}/{2}\rfloor$ and $(X^{(2)},y^{(2)})$ with sample size $n_2=n-n_1$ and estimate $\widehat{\beta}$ based on the data $(X^{(1)},y^{(1)})$ and conduct the correction in \eqref{eq: estimator Cov}  in the following form,
$
\QWh=\widehat{\beta}_{\set}^{\intercal}\widehat{\Sigma}^{(2)}_{\set,\set}\widehat{\beta}_{\set}+{2}\widehat{u}^{\intercal}(X^{(2)})^{\intercal}(y^{(2)}-X^{(2)}\widehat{\beta})/n_2
$ {with} $\widehat{\Sigma}^{(2)}=(X^{(2)})^{\intercal} X^{(2)}/n_2$ and \begin{equation}
\begin{aligned}
\widehat{u}=\arg\min \;\; u^{\intercal}\widehat{\Sigma}^{(2)} u  \quad 
{\rm s.t.} &\;\;  \max_{w\in \mathcal{C}}\left\langle w, \widehat{\Sigma}^{(2)}u -\begin{pmatrix} \widehat{\beta}_{\set}^{\intercal}\widehat{\Sigma}^{(2)}& {\bf 0}\end{pmatrix}^{\intercal}\right\rangle \leq \|\widehat{\Sigma}^{(2)}_{\set,\set}\widehat{\beta}_{\set}\|_2 \lambda_{n},
\end{aligned}
\label{eq: projection Cov split}
\end{equation}
where $\mathcal{C}$ is defined in \eqref{eq: constraint set} with $\widehat{\Sigma}_{\set,\set}$ replaced by  $\widehat{\Sigma}_{\set,\set}^{(2)}.$ 
As a result, the estimator using sample-splitting is less efficient due to the fact that only half of the data is used in  constructing the initial estimator and correcting the bias. In Theorem \ref{thm: Sigma middle} and all corresponding theoretical statement, one would then need to replace $n$ by $n/2.$ Multiple sample splitting and aggregation \citep{meinshausen2009p}, single sample splitting and cross-fitting \citep{chernozhukov2018double} or data-swapping \citep{guo2019local} are commonly used sample splitting techniques, typically improving on the inefficiency of sample splitting and reducing the dependence how the sample is actually split.

The following theorem characterizes the behaviour of the proposed estimator $\QWh$. 

\begin{Theorem} Suppose Assumptions \ref{assumption1}-\ref{assumption3} hold and $\tfrac{1}{n}\widehat{u}^{\intercal}\widehat{\Sigma}\widehat{u}$ converges in probability to a positive constant, then  $\QWh$ satisfies $\QWh-\QW=\MW+\BW$
where, as $n,p\rightarrow \infty,$
$
{\MW}\big/{({\rm V}^0_{\Sigma})^{1/2}}\rightarrow N(0,1)$ with ${\rm V}^0_{\Sigma}=\tfrac{4\sigma^2}{n}\widehat{u}^{\intercal}\widehat{\Sigma}\widehat{u}+ {\tfrac{1}{n}\E\left({\beta}_{G}^{\intercal} X_{iG} X_{iG}^{\intercal}{\beta}_{G}-{\beta}_{G}^{\intercal}{\Sigma}_{G,G}{\beta}_{G}\right)^2}
$ and
\begin{equation}
\mathbf{pr}\left\{|\BW|\geq C \big(\|\widehat{\Sigma}_{\set,\set}\widehat{\beta}_{\set}\|_2+\|\Sigma_{\set,\set}\|_2\big)\tfrac{k \log p}{n}\right\}\leq p^{-c}+\exp(-c n^{1/2}),
\label{eq: remaining}
\end{equation}
for some positive constants $C>0$ and $c>0.$ Furthermore, for any constants $\eta>0$ and $\tau>0$, under the condition $\|\beta\|_0 \leq c_1 n^{1/2}/{\log p}$ with some positive constant $c_1>0$, 
\begin{equation}
\limsup_{n,p\rightarrow \infty}\mathbf{pr}\left\{\left|\QWh-\QW\right|\geq \scale z_{1-\frac{\alpha}{2}}({\rm V}_{\Sigma})^{1/2}\right\}\leq \alpha \quad \text{with}\; {\rm V}_{\Sigma}=\tau/n+{\rm V}^0_{\Sigma}.
\label{eq: implication}
\end{equation}
\label{thm: Sigma middle}
\end{Theorem}

The above theorem establishes that the main error component $\MW$ has an asymptotic normal limit and the remaining part $\BW$ is controlled in terms of the convergence rate in \eqref{eq: remaining}. Such a decomposition is useful from the inference perspective, where \eqref{eq: implication} establishes that if the sparsity level is small enough, then the $\alpha/2$ quantile of the standardized difference $(\QWh-\QW)/({\rm V}_{\Sigma})^{1/2}$ is similar to that of the standard normal distribution, and $\BW$ is negligible in comparison to $({\rm V}_{\Sigma})^{1/2}=(\tau/n+{\rm V}^0_{\Sigma})^{1/2},$ for any constant $\tau>0$. The variance  ${{\rm V}_{\Sigma}}$ is slightly enlarged from ${\rm V}^0_{\Sigma}$ to $\tau/n+{\rm V}^0_{\Sigma}$ to quantify the uncertainty of $\MW+\BW$. Since there is no distributional result for $\BW$, an upper bound for $\BW$ would be a conservative alternative to quantify the uncertainty of $\BW$. 
The enlargement of the variance is closely related to ``super-efficiency".  The standard error $({\rm V}^0_{\Sigma})^{1/2}$ is of order ${(\|\beta_{G}\|_{2}+\|\beta_{G}\|_2^2)}/{n^{1/2}}$ and hence the variance near the null hypothesis ${\rm Q}_{\Sigma}=0$ is much faster decreasing than $1/\sqrt{n}$, which corresponds to the ``super-efficiency" phenomenon. In this case, the worst upper bound for $\BW$ is $(1+\|\beta_{G}\|_2)\cdot {k \log p}/{n},$ which can dominate $({\rm V}^0_{\Sigma})^{1/2}$ even if $k=\|\beta\|_0\leq c n^{1/2}/{\log p}.$ To overcome the challenges posed by ``super-efficiency", we enlarge the variance a bit by ${\tau}/{n}$ such that it always dominates the upper bound for $\BW$. In theory, it is sufficient to set $\tau=C {\sqrt{n}}/{k \log p}$ for a large positive constant $C>0$. However, such a selection of $\tau$ is not practical due to the unknown sparsity level and the unknown constant. In the simulation studies, we have carefully investigated the effect of $\tau$ on the proposed inference procedures. We recommend $\tau=0.5$ or $\tau=1$; see Section \ref{sec: add sim} in the supplement for the details.

Additionally, the accuracy of the test statistic depends only weakly on the group size $|G|$, in the sense that the standard deviation of the test statistic depends on $\|\beta_{G}\|_{2}$, in the order of magnitude ${(\|\beta_{G}\|_{2}+\|\beta_{G}\|_2^2)}/{n^{1/2}}$; but since $\|\beta_{G}\|_2\leq \|\beta\|_2$ and $\beta$ is sparse, the standard deviation is not always strictly increasing with a growing set $G$ and this phenomenon explains the statistical efficiency of the proposed test, especially when the test size $G$ is large. In contrast, the $\chi^2$ test proposed in \cite{mitra2016benefit,van2016chi} have the standard deviation at order of $(|G|/n)^{1/2}$, which is strictly increasing with $|G|$. This also explains their condition $|G|\ll n$.
An important feature of our proposed testing procedure is that it imposes no conditions on the group size $G$. It works for both small and large groups. 

The following theorem characterizes the estimator $\QAh$, analogously to Theorem \ref{thm: Sigma middle}.

\begin{Theorem} Suppose that Assumption \ref{assumption1}-\ref{assumption3} holds and {$\lambda_{\max}(A)$ is bounded},  then $\QAh$ satisfies $\QAh-\QA=\MA+\BA$ where as $n,p\rightarrow \infty,$
$
{\MA}\Big/{({{\rm V}^0_{\A}})^{1/2}}\rightarrow N(0,1)$ with  ${\rm V}^0_{\A}={4\sigma^2}\widehat{u}_{A}^{\intercal}\widehat{\Sigma}\widehat{u}_{A}/n$ and
\begin{equation}
\mathbf{pr}(|\BA|\gtrsim (\|A\widehat{\beta}_{\set}\|_2+\|A\|_2)\cdot{k \log p}/{n})\leq p^{-c}+\exp(-c n^{1/2})
\label{eq: remaining known}
\end{equation}
for some positive constants $C>0$ and $c>0.$  
Furthermore, for any given constants $\eta>0$ and $\tau>0$, under the condition $\|\beta\|_0 \leq c_1 n^{1/2}/{\log p}$ with some positive constant $c_1>0$,  we have 
\begin{equation}
\limsup_{n,p\rightarrow \infty}\mathbf{pr}(|\QAh-\QA|\geq z_{1-\frac{\alpha}{2}}({\rm V}_{\A})^{1/2})\leq \alpha \quad \text{with}\quad {\rm V}_{\A}={\rm V}^0_{\A}+{\tau}/{n}.
\label{eq: implication known}
\end{equation}
\label{thm: known middle}
\end{Theorem}

In contrast to Theorem \ref{thm: Sigma middle}, the main difference in Theorem \ref{thm: known middle} is  the variance level of $\MA$. In comparison, the variance of $\MW$ consists of two components, the uncertainty of estimating $\beta$ and $\Sigma$ while the variance of $\MA$ only reflects the uncertainty of estimating $\beta$.

In the following, we control the type I error of the proposed testing procedure and analyse its asymptotic power. 
We consider the following parameter space for $\theta=\left(\beta,\Sigma,\sigma \right)$,
$
\Theta\left(k\right)=\left\{\theta=\left(\beta,\Sigma,\sigma\right):
\|\beta\|_0\leq k, \; c_0\leq \lambda_{\min}\left(\Sigma\right) \leq \lambda_{\max}\left(\Sigma\right) \leq C_0,\; \sigma \leq C
\right\},
$
where $C_0>c_0>0$ and $C>0$ are positive constants. For a fixed group $G$, we define the null parameter space as 
$
\mathcal{H}_0=\left\{\theta=\left(\beta,\Sigma,\sigma\right)\in \Theta\left(k\right): \|\beta_{G}\|_2=0\right\}.
$

\begin{Corollary}Suppose that Assumption \ref{assumption1}-\ref{assumption3} holds,. For any constants $\eta>0$ and $\tau>0$, if $k \leq c n^{1/2}/\log p$ with some positive constant $c>0$, then the tests $\phi_{\Sigma}(\tau)$ in \eqref{eq: test Cov} and $\phi_{\A}(\tau)$ in \eqref{eq: test general} control the type I error, that is, $\sup_{\theta\in \mathcal{H}_0} \liminf_{n,p\rightarrow \infty}\mathbf{pr}_{\theta}\left(\phi(\tau)=1\right)\leq \alpha$ with $\phi=\phi_{\Sigma}$ or $\phi=\phi_{A}.$
\label{cor: 1}
\end{Corollary}
We present the power result in the following corollary and define the local alternative hypothesis parameter space as 
$
\mathcal{H}_{1}(A, \delta)=\left\{\theta=\left(\beta,\Sigma,\sigma\right)\in \Theta\left(k\right): \beta_{G}^{\intercal}A\beta_{G}={\delta}\right\}.
$
\begin{Corollary}
Suppose that the assumptions of Corollary \ref{cor: 1} hold. For any $\theta\in \mathcal{H}_{1}(\Sigma_{G,G},\delta(t))$ with $\delta(t)=((1+2\eta)z_{1-\alpha}+t)(\V_{\Sigma})^{1/2},$   
the test $\phi_{\Sigma}(\tau)$ in \eqref{eq: test Cov} has the asymptotic power
$
\liminf_{n,p\rightarrow \infty} \mathbf{pr}_{\theta}\left(\phi_{\Sigma}(\tau)=1\right)\geq 1-\Phi(-t), 
$
where ${\rm \V}_{\Sigma}$ is defined in \eqref{eq: implication} and $\Phi(\cdot)$ is the quantile function of standard normal distribution.
For any $\theta\in \mathcal{H}_{1}(A, \delta(t))$ with $\delta(t)=((1+2\eta)z_{1-\alpha}+t)(\V_{\A})^{1/2},$   
the test $\phi_{\A}(\tau)$ in \eqref{eq: test general} has the asymptotic power,
$
\liminf_{n,p\rightarrow \infty} \mathbf{pr}_{\theta}\left(\phi_{\Sigma}(\tau)=1\right)\geq 1-\Phi(-t) 
$
where ${\rm \V}_{\A}$ is defined in \eqref{eq: implication known}.
\label{cor: 2}
\end{Corollary}
As a remark,  the separation parameter of the defined local alternative space $\delta(t)=((1+2\eta)z_{1-\alpha}+t)(\V_{\Sigma})^{1/2}$ is of the order $({\tau}^{1/2}+\|\beta_{G}\|_{2}+\|\beta_{G}\|_2^2)/n^{1/2}$ and the proposed test $\phi_{\Sigma}$ has power converging to 1 as long as $t\rightarrow\infty$. 
It is interesting to make a more technical comparison with the maximum test with or without the bootstrapped version. Consider a challenging setting for the significance test, where $\beta_i\in \{0,c_{\beta}({\log |G|}/{n})^{1/2}\}$ for $i\in \set$, the signal is relatively dense with $k=\|\beta\|_0=c \sqrt{n}/\log p$ for some constant $c>0$ and the group size $|G|\asymp p$.
It is known that the maximum test has nontrivial power for sufficiently large $c_{\beta}$. Let $k_{G}=|\{i\in \set: \beta_i \neq 0\}|$; when $|G|$ gets larger, we have $k_{\set} \approx k$ and $\|\beta_{\set}\|_2^2 =c_{\beta}^2 k_{\set} \log |G|/n \approx c_{\beta}^2 c/n^{1/2}\cdot (\log |G|/\log p)$. Corollary \ref{cor: 2} implies that the proposed test has nontrivial power for a sufficiently large $c_{\beta}$. Hence, the maximum test and the proposed test have the same boundary in terms of convergence rate and the difference is in the order of a constant. We have further examined the finite sample numerical performance in such scenarios and observe that the proposed test can be much more powerful than the maximum test. See Section \ref{sec: sim dense} for details.

\section{Statistical Applications}
\label{sec: application}

\subsection{Hierarchical testing}
\label{sec: hier appl}
It is often too ambitious to detect significant single variables,
in particular in presence of high correlation or near collinearity among the
variables. On the other hand, a group is easier to be
detected as significant, especially in presence of strong correlation. Hierarchical sequential testing is a powerful method to go through a sequence of significance tests, from larger groups to smaller ones. Thanks to its sequential nature, it automatically adapts to the strength of the signal and in relation to correlation among the variables, and it does not need any pre-specification of the size of the groups to be tested. It is typically much more powerful than performing Bonferroni-Holm adjustment over the entire number of hypotheses under consideration. 
Such hierarchical procedures have been proposed in general by \cite{meinshausen2008hierarchical}, tailored for high-dimensional regression by \cite{manpb16}, \cite{renaux2020hierarchical} and used and validated in applications in \cite{buzduganetal16} and \cite{klasen2016multi}. 

A particular hierarchical testing scheme is described in Supplement \ref{appendix: hier appl}. It requires as input a hierarchical tree of groups of variables, often taken as a hierarchical cluster tree based on the covariates only. Starting at the top of the tree, corresponding to the group of all variables $\{1,\ldots ,p\}$, our proposed group test is used and if found to be significant, we proceed by testing more refined groups $G \subseteq \{1,\ldots, p\}$ further down the tree. As output we obtain a set of significant (disjoint) groups such that the familywise error rate is controlled.
Our new group significance test for hierarchical sequential testing $H_{0,G}: \beta_G = 0$ for many different $G$ is perfectly tailored for such hierarchical applications as it is computationally fast and has good power properties. In practice, we recommend using $\phi_{\Sigma}(\tau)$ in \eqref{eq: test Cov} with $\tau=0.5$ to test $H_{0,G}.$

\subsection{Testing interaction and detection of effect heterogeneity}
\label{sec: interaction}
The proposed significance test is useful in testing the existence of interaction, which itself is an important statistical problem. We focus on the interaction model 
$
y_i=X_{i\cdot}^{\intercal}\beta+D_i \cdot X_{i\cdot}^{\intercal}\gamma+\epsilon_i
$
and  re-express the model as 
$y_i=W_{i\cdot}^{\intercal}\eta+\epsilon_i$
with $W_i=(D_i\cdot X^{\intercal}_{i\cdot},X^{\intercal}_{i\cdot})^{\intercal}$ and $\eta=(\gamma^{\intercal},\beta^{\intercal})^{\intercal}$. {We adopt the convention that $X_{i1}=1$ and then the detection of interaction terms between $D_i$ and $X_{i,-1}$ is reduced to the group significance test  
$H_0: \eta_{G}=0$ for $G=\{2,\ldots,p\}.$} The detection of heterogeneous treatment effect can be viewed as a special case of testing the interaction term. 
If $D_{i}$ in the interaction model is taken as a binary variable, where $D_i=0$ or $1$ denotes that the subject belongs to the control group or the treatment group, respectively, then this specific test for interaction amounts to testing whether the treatment effect is heterogeneous. In a very similar way, if $D_{i}$ takes two values where $D_{i}=1$ and $D_{i}=2$ represent the subject is receiving treatment 1 and 2, respectively, then the test of interaction is for testing whether the difference between two treatment effects is heterogeneous. The current developed method of detecting heterogeneous treatment effects is definitely not restricted to the case of a binary treatment. It can be applied to basically any type of treatment variables, such as count, categorical or continuous variables.

\subsection{Local heritability}
\label{sec: local herit}
Local heritability is defined as a measure of the partial variance explained by a given set of genetic variables. In contrast to the (global) heritability, the local heritability is more informative as it describes the variability explained by a pre-specified set of genetic variants and takes the global heritability as one special case. Assuming the regression model \eqref{eq: linear model}, the local heritability can be represented by the quantities, $\beta_{G}^{\intercal}\Sigma_{G,G} \beta_{G}$ and $\|\beta_{G}\|_2^2$, where $G$ denotes the index set of interest. The following corollary establishes the coverage properties of the proposed confidence intervals for two measures of local heritability, $\beta_{G}^{\intercal}\Sigma_{G,G} \beta_{G}$ and $\|\beta_{G}\|_2^2$.

\begin{Corollary}
Suppose that Assumption \ref{assumption1}-\ref{assumption3} holds and $\tau>0$ is a positive constant, then 
\begin{equation*}
\liminf_{n,p\rightarrow \infty}\mathbf{pr}(\QW\in {\rm CI}_{\Sigma}(\tau))\geq 1-\alpha, \quad \liminf_{n,p\rightarrow \infty}\mathbf{pr}(\QA\in {\rm CI}_{A}(\tau))\geq 1-\alpha,
\end{equation*}
where  ${\rm CI}_{\Sigma}(\tau)$ and ${\rm CI}_{A}(\tau)$ are defined in \eqref{eq: CI Cov} and \eqref{eq: CI general}, respectively.
\label{cor: local heritability}
\end{Corollary}

\vspace{-10pt}
\section{Simulation Study and Real Data Application}
\label{sec: simulation}
\subsection{General set up}
Throughout the simulation study, we consider high dimensional linear models
$
y_{i}=\sum_{j=1}^{p}X_{ij}\beta_j+\epsilon_{i}$ for  $i = 1, \ldots, n,$
with  $p=500$. We generate the covariates following $X_{i\cdot}\sim N({\bf 0},\Sigma)$ and the error $\epsilon_{i} \sim N(0,1),$ both being independent of each other and independent and identically distributed over the indices $i$. The results are calculated based on 500 simulation runs.

We take $\widehat{\beta}$ as the Lasso estimator \citep{tibshirani1996regression}, which is computed using the R-package \texttt{cv.glmnet} \citep{glmnet} with the tuning parameter $\lambda$ chosen by cross-validation. 
 For the construction of the projection direction in \eqref{eq: projection Cov}, we first solve its dual problem 
\begin{equation*}
\widehat{v}=\argmin_{v \in \R^{p+1}} \frac{1}{4}v^{\intercal} H^{\intercal}\widehat{\Sigma}H v+b^{\intercal} H v+\lambda \|v\|_1\;\text{with}\; H=\left[\begin{matrix} b, \mathbb{I}_{p\times p} \end{matrix} \right], \; b=\left(\tfrac{\widehat{\beta}_{\set}^{\intercal}\widehat{\Sigma}_{\set,\set}}{\|\widehat{\Sigma}_{\set,\set}\widehat{\beta}_{\set}\|_2}, {\bf 0}^{\intercal}\right)^{\intercal}
\label{eq: dual problem}
\end{equation*}
where we adopt the notation $0/0=0$. We then construct the direction as $\widehat{u}=-(\widehat{v}_{-1}+\widehat{v}_1 b)/2.$ When $H^{\intercal}\widehat{\Sigma}H$ is singular and $\lambda$ is close to zero, then dual problem is unbounded from below. Hence, the tuning parameter $\lambda$ is chosen as the smallest $\lambda>0$ such that the dual problem has a bounded optimal value. The algorithm implementation with tuning parameter selection can be found at \texttt{https://statistics.rutgers.edu/home/zijguo/Software.html}.

We consider four specific tests, $\phi_{\rm I}(0.5),\phi_{\rm I}(1),\phi_{\Sigma}(0.5),\phi_{\Sigma}(1)$ where $\phi_{\Sigma}(0.5),\phi_{\Sigma}(1)$ are defined in \eqref{eq: test Cov} with $\tau=0.5$ and $\tau=1$, respectively and $\phi_{\rm I}(0.5),\phi_{\rm I}(1)$ are defined in \eqref{eq: test Cov} (by taking $A={\rm I}$) with $\tau=0$ and $\tau=1$, respectively. We set $\tau=0.5$ or $\tau=1$ thereby providing a conservative upper bound for the bias component. We explore how the value $\tau$ affects the performance of the proposed methods in Section \ref{sec: add sim} in the supplement.

 The proposed method is compared with two alternative procedures, the maximum test based on the debiased estimator proposed in \cite{javanmard2014confidence}, shorthanded as \texttt{FD} (Fast Debiased) and the maximum test based on the debiased estimator proposed in \cite{van2014asymptotically}, shorthanded as \texttt{hdi}. 
Specifically, we produce the \texttt{FD} debiased estimators $\{\widehat{\beta}_{j}^{\rm \texttt{FD}}\}_{j = 1, \ldots p}$ by the online code of \cite{javanmard2014confidence} and the \texttt{hdi} estimator $\{\widehat{\beta}_{j}^{\rm \texttt{hdi}}\}_{j = 1, \ldots p}$  by using the R package \texttt{\texttt{hdi}} \citep{hdipackage}. The additional products of these implemented algorithms include the corresponding covariance matrix, denoted as ${\rm cov}(\widehat{\beta}^{\rm \texttt{FD}})\in \R^{p\times p}$ and ${\rm cov}(\widehat{\beta}^{\rm \texttt{hdi}})\in \R^{p\times p},$ respectively. 
For the maximum test for group $G,$ we sample independent and identically distributed copies $Z_1,\ldots, Z_{10,000}\in \R^{|G|}$ following $N\big({\bf 0}, {\rm cov}(\widehat{\beta}^{\rm \texttt{FD}})_{G\times G}\big)$ and calculate $q^{\rm \texttt{FD}}_{\alpha}$ as the empirical $1-\alpha$ quantile of $\max_{j\in G}|Z_{1,j}|, 
\ldots, \max_{j\in G}|Z_{10,000,j}|.$ We similarly define $q^{\rm \texttt{hdi}}_{\alpha}$ by replacing ${\rm cov}(\widehat{\beta}^{\rm \texttt{FD}})_{G\times G}$ with ${\rm cov}(\widehat{\beta}^{\rm \texttt{hdi}})_{G\times G}$. Then we define the following tests for group significance 
$
\phi_{\rm \texttt{FD}}=\mathbf{1}(\max_{j\in G}|\widehat{\beta}_{j}^{\rm \texttt{FD}}|\geq q^{\rm \texttt{FD}}_{\alpha})$ and $\phi_{\rm \texttt{hdi}}=\mathbf{1}(\max_{j\in G}|\widehat{\beta}_{j}^{\rm \texttt{hdi}}|\geq q^{\rm \texttt{hdi}}_{\alpha}).
$

We shall compare $\phi_{\rm I}(0.5),\phi_{\rm I}(1),\phi_{\Sigma}(0.5),\phi_{\Sigma}(1)$ and $\phi_{\rm \texttt{FD}},\phi_{\rm \texttt{hdi}}$ in two settings, dense alternatives (Section \ref{sec: sim dense}) and high correlation (Section \ref{sec: sim high-cor}).

\subsection{Dense alternatives}
\label{sec: sim dense}
In this section, we consider the setting where the regression vector is relatively dense but with small non-zero coefficients, as this is a challenging scenario for detecting the signals. We generate the regression vector $\beta$ as $\beta_j=\delta$ for $25\leq j\leq 50$  and $\beta_j=0$ otherwise and generate the covariance matrix $\Sigma_{ij}=0.6^{|i-j|}$ for $1\leq i,j\leq 500$.  We consider the group significance test,
$
H_{0,G}: \beta_{i}=0 \; \text{for} \; i\in G, $ with  $G=\{30,31,\ldots,200\}. 
$ We vary the signal strength parameter $\delta$ over $\{0, 0.04,0.06\}$ and the sample size $n$ over $\{250,350,500\}$. 

\begin{table}[ht]
\centering
\begin{tabular}{|rr|rr|rr|rr|}
  \hline
$\delta$ &$n$& $\phi_{\rm I}(0.5)$ & $\phi_{\rm I}(1)$  & $\phi_{\Sigma}(0.5)$ & $\phi_{\Sigma}(1)$ & $\phi_{\rm \texttt{FD}}$& $\phi_{\rm \texttt{hdi}}$ \\ 
  \hline
\multirow{ 4}{*}{0} &250& 0.002 & 0.000 & 0.016 & 0.004 & 0.112 & 0.044 \\ 
                             &350 & 0.002 & 0.002 & 0.014 & 0.006 & 0.086 & 0.042 \\ 
                              &500& 0.006 & 0.000 & 0.006 & 0.000 & 0.078 & 0.048 \\ 
  \hline
\multirow{ 4}{*}{0.04} &250&0.182 & 0.040 & 0.568 & 0.350 & 0.226 & 0.084 \\  
                                   &350 & 0.338 & 0.062 & 0.750 & 0.504 & 0.184 & 0.106 \\ 
                                   &500 & 0.518 & 0.170 & 0.928 & 0.732 & 0.128 & 0.112 \\ 
  \hline
\multirow{ 4}{*}{0.06} &250& 0.770 & 0.400 & 0.978 & 0.938 & 0.344 & 0.162 \\ 
                                  &350 & 0.928 & 0.650 & 0.998 & 0.996 & 0.316 & 0.192 \\ 
                                  &500& 0.998 & 0.902 & 1.000 & 1.000 & 0.252 & 0.272 \\ 
  \hline
\end{tabular}
\caption{Empirical Rejection Rate (ERR) for the Dense Alternative scenario (5\% significance level). We report the ERR for six different tests $\phi_{\rm I}(0.5), \phi_{\rm I}(1), \phi_{\Sigma}(0.5), \phi_{\Sigma}(1),\phi_{\rm \texttt{FD}}$ and $\phi_{\rm \texttt{hdi}},$ where ERR denotes the proportion of rejected hypothesis among the total $500$ simulations.}
\label{tab: decision dense}
\end{table}

Table \ref{tab: decision dense} summarizes the hypothesis testing results. 
For $\delta=0$, the empirical detection rate is an empirical measure of the type I error; For $\delta \neq 0$,  the empirical detection rate is an empirical measure of the power.  The proposed procedures $\phi_{\rm I}(0.5), \phi_{\rm I}(1), \phi_{\Sigma}(0.5)$ and $\phi_{\Sigma}(1)$ control the type I error. As comparison, the maximum test $\phi_{\rm \texttt{hdi}}$ controls the type I error while the other maximum test  $\phi_{\rm \texttt{FD}}$ does not reliably control the type I error. To compare the power, we observe that $\phi_{\Sigma}(0.5)$ and $\phi_{\Sigma}(1)$ are in general more powerful than the corresponding $\phi_{\rm I}(0.5)$ and $\phi_{\rm I}(1)$. Across all settings, the power of both $\phi_{\rm \texttt{hdi}}$ and $\phi_{\rm \texttt{FD}}$ are lower than the proposed $\phi_{\Sigma}(0.5),\phi_{\Sigma}(1).$ In most settings, the power of both $\phi_{\rm \texttt{hdi}}$ and $\phi_{\rm \texttt{FD}}$ are much lower than $\phi_{\rm I}(0.5)$.  An interesting observation is that, although the proposed testing procedures $\phi_{\Sigma}(1)$ and $\phi_{\rm I}(1)$  control the type I error in a conservative sense, they still achieve a higher power than the existing maximum tests. 

We report the computational time (averaged  over 50 simulations) of $\phi_{\rm \texttt{hdi}}$, $\phi_{\rm \texttt{FD}}$ and $\phi_{\rm I}$ and $\phi_{\Sigma}$ in Table \ref{tab: time}. The proposed methods $\phi_{\rm I}$ and $\phi_{\Sigma}$ are computationally efficient as they correct the bias all at once while the bias correction step of $\phi_{\rm \texttt{hdi}}$ or $\phi_{\rm \texttt{FD}}$ requires the implementation of $|G|=171$ penalized regression in the dimension of $p=500$.  
\begin{table}[htp!]
\centering
\begin{tabular}{|rr|r|r|r|r|}
  \hline
  $\delta$ &n & $\phi_{\rm I}$ & $\phi_{\Sigma}$ & {\rm \texttt{FD}} & {\rm \texttt{hdi}} \\ 
  \hline
\multirow{ 3}{*}{0} &250  & 10.82 & 10.93 & 74.37 & 274.42 \\ 
  & 350 & 17.35 & 22.60 & 65.07 & 297.36 \\ 
  & 500 & 37.79 & 35.30 & 88.76 & 2202.34 \\ 
  \hline
 \multirow{ 3}{*}{0.04} &250 &  17.57 & 23.30 & 78.09 & 283.82 \\ 
  &350 & 37.50 & 48.78 & 64.58 & 296.27 \\ 
  &500 & 58.08 & 77.87 & 89.13 & 2226.97 \\ 
  \hline
  \multirow{ 3}{*}{0.06} &250 & 15.73 & 24.20 & 72.81 & 269.28 \\ 
  &350 & 42.00 & 67.35 & 113.01 & 475.94 \\ 
  &500 & 49.74 & 76.06 & 89.66 & 2303.65 \\ 
   \hline
\end{tabular}
\caption{Average computing time (in seconds) over $50$ simulation for the Dense Alternative scenario.}
\label{tab: time}
\end{table}

In Figure \ref{fig: dense-det} in the supplement, we report the ERR for other choices of $\tau$ and observe that for $\tau=0$, the testing procedures $\phi_{\rm I}(0)$ and $\phi_{\Sigma}(0)$ do not reliably control the type I error while for $\tau\geq 0.5$ they do. This matches with the theoretical results in Corollary \ref{cor: 1}, where a positive constant $\tau>0$ is needed to address the super-efficiency and control the type I error. In Figure \ref{fig: dense-cov} in the supplement, we have further explored the coverage properties for the proposed confidence intervals ${\rm CI}_{\rm I}(\tau)$ for $\|\beta_{G}\|_2^2$ and  ${\rm CI}_{\Sigma}(\tau)$ for $\beta_{G}^{\intercal}\Sigma_{G,G}\beta_{G}$: ${\rm CI}_{\rm I}(\tau=0)$ and ${\rm CI}_{\Sigma}(\tau=0)$ are not guaranteed to have coverage while ${\rm CI}_{\rm I}(\tau)$ and ${\rm CI}_{\Sigma}(\tau)$, for $\tau\geq 0.5$, nearly achieve the desired coverage levels in most settings. 

We have examined the performance of the data-splitting version of the algorithm described in \eqref{eq: projection Cov split}. In Figures \ref{fig: dense-det-split} and \ref{fig: dense-cov-split} in the supplement, we observe that the testing procedures and confidence intervals with data-splitting are worse than those using the full data {(except for type I error control, which reliably holds with sample splitting as well). However, with a larger sample size, the testing procedures  achieve reasonable power and the confidence intervals attain the coverage level.} The sample splitting is only introduced to facilitate the technical proof and the procedures using the full data work well in practical settings.

We have also explored the testing and coverage properties over different sparsity levels and report the results in Figures \ref{fig: dense-det-sparser} and \ref{fig: dense-cov-sparser} in the supplement.

\subsection{Highly correlated covariates}
\label{sec: sim high-cor}
Here, we consider the setting where the regression vector is relatively sparse but a few variables are highly correlated. We generate the regression vector $\beta$ as $\beta_1=\beta_3=\delta$ and $\beta_j=0$ for $j\neq 1,3$ and we generate the covariance matrix as follows: $\Sigma_{ij}=0.8$ for $1\leq i\neq j\leq 5$ and $\Sigma_{ij}=0.6^{|i-j|}$, otherwise. 
There exists high correlations among the first five variables, where the pairwise correlation is $0.8$ inside this group of five variables. In contrast to the previous simulation setting in Section \ref{sec: sim dense}, we do not generate a large number of non-zero entries in the regression coefficient but only assign the first and third coefficients to be possibly non-zero.  We test the group hypothesis generated by the first five regression coefficients,
$
H_{0,G}: \beta_{i}=0 \; \text{for} \; i\in G, \; \text{where} \; G=\{1,2,\ldots,5\}. 
$
We vary the signal strength parameter $\delta$ over $\{0,0.2,0.3\}$ and the sample size $n$ over $\{250,350,500\}$. 

As reported in Table \ref{tab: decision high}, the proposed testing $\phi_{\rm I}(0.5), \phi_{\rm I}(1), \phi_{\Sigma}(0.5), \phi_{\Sigma}(1)$ and the maximum test procedure $\phi_{\rm \texttt{hdi}}$ control the type I error while $\phi_{\rm \texttt{FD}}$ barely controls it.
Regarding the power, $\phi_{\rm \texttt{hdi}}$ and $\phi_{\rm \texttt{FD}}$ are better for $\delta=0.2$ while our proposed testing procedures $\phi_{\Sigma}(0.5)$ and $\phi_{\Sigma}(1)$ are comparable to $\phi_{\rm \texttt{hdi}}$ and $\phi_{\rm \texttt{FD}}$ when $\delta$ reaches $0.3$.

Seemingly, our proposed procedures $\phi_{\Sigma}$ and $\phi_{\rm I}$ do not perform better than the maximum test  $\phi_{\rm \texttt{hdi}}$ and $\phi_{\rm \texttt{FD}}.$ The performance of the latter two for testing is in sharp contrast to the individual coverage. We shall emphasize that the individual coverage properties related to the maximum test  $\phi_{\rm \texttt{hdi}}$ and $\phi_{\rm \texttt{FD}}$ are not guaranteed although this is not visible in Table \ref{tab: decision high}. Specifically, since we are testing $\beta_i=0$ for $i=1,2,3,4,5$, we can look at the coverage properties of these two proposed tests in terms of $\beta_i$. As reported in Table \ref{tab: IC high}, for $\delta \neq 0,$ we have observed that the coordinate-wise coverage properties are not guaranteed due to the high correlation among the first five variables. The reason for this phenomenon is that the coverage for an individual coordinate $\beta_j$ requires a decoupling between the $j$-th and all other variables and if there exists high correlations, this decoupling step is difficult to be conducted accurately. In contrast, even though the first five variables are highly correlated, the constructed confidence intervals $ {\rm CI}_{\rm I}(\tau=0.5), {\rm CI}_{\rm I}(\tau=1), {\rm CI}_{\Sigma}(\tau=0.5)$ and ${\rm CI}_{\Sigma}(\tau=1)$ achieve the $95\%$ coverage. This is reported in Table \ref{tab: coverage high}. Our proposed testing procedure is more robust to high correlations inside the testing group as the whole group is tested as a unit instead of decoupling variables inside the testing group.

We explore the effect of $\tau$ in the supplement and report the dependence of the proposed testing procedure on $\tau$ in Figure \ref{fig: HC-det} and the dependence of the coverage properties on $\tau$ in Figure \ref{fig: HC-cov}. The phenomenon is similar to the dense alternative setting: the proposed methods are reliable for $\tau\geq 0.5.$

\begin{table}[ht]
\centering
\begin{tabular}{|rr|rr|rr|rr|}
  \hline
$\delta$ &$n$& $\phi_{\rm I}(0.5)$ & $\phi_{\rm I}(1)$ & $\phi_{\Sigma}(0.5)$ & $\phi_{\Sigma}(1)$ & $\phi_{\rm \texttt{FD}}$& $\phi_{\rm \texttt{hdi}}$ \\ 
  \hline
\multirow{ 3}{*}{0} &250& 0.000 & 0.000 & 0.000& 0.000 & 0.070 & 0.036 \\ 
                             &350 & 0.000 & 0.000 & 0.000 & 0.000 & 0.082 & 0.062 \\ 
                             &500& 0.000 & 0.000 & 0.000 & 0.000 & 0.082 & 0.056 \\ 
 \hline
\multirow{ 3}{*}{0.2} &250& 0.194& 0.112 & 0.674 &0.414& 0.998 & 0.936 \\ 
                                &350 & 0.196 & 0.124 & 0.878& 0.692 & 1.000 & 0.972 \\ 
                                &500 & 0.272 & 0.166 & 0.984&0.924 & 1.000 & 0.994 \\ 
 \hline
\multirow{ 3}{*}{0.3} &250& 0.472 & 0.384 & 1.000 & 0.998 & 1.000 & 1.000 \\ 
                                &350 & 0.462 & 0.434 & 1.000 & 1.000 & 1.000 & 1.000 \\ 
                                &500 & 0.518 & 0.498 & 1.000 & 1.000 & 1.000 & 1.000 \\ 
  \hline
\end{tabular}
\caption{Empirical Rejection Rate for the Highly Correlated scenario (5\% significance level). We report the ERR for six different tests $\phi_{\rm I}(0.5), \phi_{\rm I}(1), \phi_{\Sigma}(0.5), \phi_{\Sigma}(1),\phi_{\rm \texttt{FD}}$ and $\phi_{\rm \texttt{hdi}},$ where ERR denotes the proportion of rejected hypothesis among the total $500$ simulations.}
\label{tab: decision high}
\end{table}

\begin{table}[ht]
\centering
\begin{tabular}{|rr|rrrrr|rrrrr|}
  \hline
&& \multicolumn{5}{c}{${\rm CI}_{\rm \texttt{FD}}$}\vline & \multicolumn{5}{c}{${\rm CI}_{\rm \texttt{hdi}}$}\vline\\
\hline
$\delta$ & $n$  & $\beta_1$& $\beta_2$ & $\beta_3$ & $\beta_4$ & $\beta_5$ & $\beta_1$& $\beta_2$ & $\beta_3$ & $\beta_4$ & $\beta_5$ \\ 
  \hline
\multirow{ 3}{*}{0} &250& 0.972 & 0.968 & 0.970 & 0.976 & 0.976 & 0.952 & 0.950 & 0.944 & 0.950 & 0.946 \\ 
  &350& 0.968 & 0.972 & 0.962 & 0.970 & 0.968 & 0.942 & 0.942 & 0.932 & 0.966 & 0.948 \\ 
 & 500 & 0.974 & 0.972 & 0.964 & 0.970 & 0.982 & 0.950 & 0.936 & 0.956 & 0.950 & 0.956 \\ 
%
\hline
\multirow{ 3}{*}{0.2} &250& 0.400 & 0.714 & 0.418 & 0.720 & 0.758 & 0.864 & 0.798 & 0.910 & 0.828 & 0.268 \\ 
  &350& 0.464 & 0.696 & 0.414 & 0.722 & 0.680 & 0.910 & 0.822 & 0.922 & 0.844 & 0.234 \\ 
  & 500 & 0.424 & 0.702 & 0.408 & 0.686 & 0.674 & 0.876 & 0.860 & 0.916 & 0.842 & 0.298 \\ 
   \hline
\multirow{ 3}{*}{0.3} &250& 0.430 & 0.724 & 0.426 & 0.720 & 0.740 & 0.870 & 0.808 & 0.890 & 0.818 & 0.218 \\ 
  &350 & 0.386 & 0.732 & 0.432 & 0.682 & 0.720 & 0.832 & 0.836 & 0.904 & 0.836 & 0.258 \\ 
  & 500 & 0.422 & 0.692 & 0.426 & 0.694 & 0.686 & 0.860 & 0.856 & 0.900 & 0.854 & 0.280 \\ 
   \hline
\end{tabular}
\caption{Empirical Coverage for first five regression coefficients for Highly Correlated scenario (95\% nominal coverage). The numbers under ${\rm CI}_{\rm \texttt{FD}}$ represent the empirical coverage for $\beta_j$ ($1\leq j\leq 5$) using the method proposed in \cite{javanmard2014confidence} and the numbers under ${\rm CI}_{\rm \texttt{hdi}}$ represent the empirical coverage for $\beta_j$ ($1\leq j\leq 5$) using the method proposed in \cite{van2014asymptotically}.}
\label{tab: IC high}
\end{table}

\begin{table}[ht]
\centering
\begin{tabular}{|rr|rr|rr|rr|}
  \hline
$\delta$ &$n$&${\rm CI}_{\rm I}(\tau=0.5)$& ${\rm CI}_{\rm I}(\tau=1)$ & ${\rm CI}_{\Sigma}(\tau=0.5)$& ${\rm CI}_{\Sigma}(\tau=1)$ \\ 
  \hline
\multirow{ 3}{*}{0} &250& 1.000 & 1.000 & 1.000 & 1.000 \\ 
                             &350& 1.000 & 1.000 & 1.000 & 1.000 \\ 
                            & 500 & 0.988& 0.988 & 0.986 & 0.986 \\ 
\hline
\multirow{ 3}{*}{0.2} &250& 0.992 & 0.992 & 0.942 & 0.994\\ 
                                &350 & 0.998 & 0.998 & 0.972 & 0.998 \\ 
                                & 500& 0.990 & 0.992 & 0.970 & 1.000 \\ 
\hline
\multirow{ 3}{*}{0.3} &250& 0.970 & 0.986 & 0.916 & 0.962 \\ 
                                &350 & 0.966 & 0.986 & 0.900 & 0.954 \\ 
                               & 500 & 0.966 & 0.978 & 0.954 & 0.972 \\ 
\hline
\end{tabular}
\caption{Empirical Coverage for the Highly Correlated scenario (95\% nominal coverage). We report the empirical coverage of ${\rm CI}_{\rm I}(\tau=0.5)$ and ${\rm CI}_{\rm I}(\tau=1)$ for $\|\beta_{G}\|_2^2$ and the empirical coverage of ${\rm CI}_{\Sigma}(\tau=0.5)$ and ${\rm CI}_{\Sigma}(\tau=1)$ for $\beta^{\intercal}_{G}\Sigma_{G,G}\beta_{G}.$}
\label{tab: coverage high}
\end{table}

\subsection{Hierarchical testing}
\label{sec: sim hier}

We simulate data under two settings which differ in the set of active covariates $S={\rm supp}(\beta)$ and $\Sigma.$
In both cases, $\Sigma$ is block diagonal. 
We fix $p=500$, $|S|=\|\beta\|_0=10$, and $\beta_{j} = 1$ for $j \in S$ 
and vary the number of observations $n$ between 
$100, 200, 300, 500,\mbox{ and } 800$.

In \textit{setting 1}, the first 20 covariates have high correlations within small blocks of size 2. The covariance 
matrix $\Sigma$ has 1's on the diagonal, 
$\Sigma_{i, i + 1} = \Sigma_{i + 1, i} = 0.7$ 
for $i = 1, 3, 5, \ldots, 19$, and 0's otherwise. The set of 
active covariates is $S = \{1, 3, 5, \ldots, 19\}$. 

In \textit{setting 2}, there are ten blocks each corresponding to 50 covariates 
that have a high pairwise correlation of 0.7. The covariance matrix 
$\Sigma$ has 1's on the diagonal, $\Sigma_{B_l} = 0.7$ for 
$B_l = \big\{(i,j): i \neq j \mbox{ and } i, j \in \{l, l + 1, \ldots, l + 50\}\big\}$ 
for $l = 1, 51, 101, \ldots$ 451, and 0's otherwise. 
The set of active covariates is $S = \{1, 51, 101, \ldots, 451\}$. 

For every simulation run, a hierarchical cluster tree is estimated 
using $1 - (\text{empirical correlation})^2$ as
dissimilarity measure and average linkage.
The hierarchical procedure with our proposed group testing method 
$\phi_{\Sigma}(\tau=1)$ 
performs testing top-down through this tree.

We do not consider a single group but instead, we aim with hierarchical testing to find as many significant groups as possible.
We use a modified version of the power to measure the performance of the 
hierarchical procedure because groups of variable sizes are returned. 
The adaptive power is defined by  
$
\mbox{Power}_{\mbox{\footnotesize{adap}}} = ({1}/{|S|})\cdot \sum\limits_{C \, \in \, \mbox{\footnotesize{MTD}}} {1}/{|C|}
$
where MTD stands for Minimal True Detections, i.e., there is no significant 
subgroup (``Minimal''), the group has to be significant (``Detection''), and 
the group contains at least one active variable (``True''); see \cite{manpb16}.  

The results are reported in 
Table \ref{tab.hierGroup}.
The hierarchical procedure performs very well for setting 1 with the adaptive 
power around 0.9 till 1.0 and the 
procedure finds 10 significant groups of average size 1.0 till 1.2 for all values 
of $n$ except $n = 100$. 
Setting 2 is much harder because the 10 active covariates are each 
highly correlated with 49 non-active covariates. It is difficult to distinguish 
the active variables from the correlated ones and hence, the procedure 
stops further up in the tree resulting in larger significant groups
and smaller adaptive power compared 
to setting 1. 
Note that the usual measure of power (where a significant group is counted
as true if at least one active covariate is in it) 
is close to or even 1 for both settings. 
The familywise error rate is well controlled for both settings. 

\begin{table}[ht]
\centering
\begin{tabular}{|lr|rrr|rrr|}
  \hline
Setting & $n$  & FWER & power & 
adaptive power & avg number & avg size & median size \\ 
  \hline
  1 & 100 & 0.038 &  0.923 & 
  0.692 & 9.3 & 4.6 & 1.0 \\ 
  1 & 200 & 0.004 &  1.000 & 
  0.947 & 10.0 & 1.2 & 1.0 \\ 
  1 & 300 & 0.002 &  1.000 & 
  0.898 & 10.0 & 1.2 & 1.0 \\ 
  1 & 500 & 0.000 &  1.000 & 
  0.985 & 10.0 & 1.0 & 1.0 \\ 
  1 & 800 & 0.000 &  1.000 & 
  1.000 & 10.0 & 1.0 & 1.0 \\ 
  \hline
  2 & 100 & 0.082 &  0.959 & 
  0.185 & 9.7 & 32.8 & 40.5 \\ 
  2 & 200 & 0.006 &  1.000 & 
  0.175 & 10.0 & 30.3 & 39.5 \\ 
  2 & 300 & 0.002 &  1.000 & 
  0.428 & 10.0 & 19.2 & 16.0 \\ 
  2 & 500 & 0.002 &  1.000 & 
  0.478 & 10.0 & 18.8 & 15.8 \\ 
  2 & 800 & 0.000 &  1.000 & 
  0.766 & 10.0 & 8.6 & 1.0 \\ 
   \hline
\end{tabular}
\caption{Results from the simulation study of the hierarchical procedure (FWER level $5\%$). The last six columns are familywise error rate (FWER), power, adaptive power, average number, average size, and median size of the significant groups.} 
\label{tab.hierGroup}
\end{table}

\vspace{-10pt}

\subsection{Real Data Analysis for Yeast Colony Growth} 
\label{sec: colony}

\cite{bloom2013finding} performed a genome-wide association 
study of 46 quantitative traits to investigate the sources of missing 
heritability. The authors crossbred 1,008 yeast Saccharomyces cerevisiae 
segregates from a laboratory strain and a wine strain and measured 11,623 
genotype markers which they reduced to 4,410 markers that show less correlation. 
\cite{bloom2013finding} processed the data such that the covariates 
encode from which of the two strains a given genotype was passed on. This is 
encoded using the values $1$ and $-1$.  
Each crossbred was exposed to 46 different conditions like different 
temperatures, pH values, carbon sources, additional metal ions, and small 
molecules. The traits of interest are the end-point colony size normalized 
by the colony size on control medium, see \cite{bloom2013finding} for further 
details.  

We use this data set to illustrate the hierarchical procedure with our proposed
group testing method $\phi_{\Sigma}(\tau=1)$. 
We consider each trait separately resulting in 46 different regression problems. 
The hierarchical procedure goes top-down through a hierarchical cluster
tree which was estimated using $1 - (\text{empirical correlation})^2$ as
dissimilarity measure and average linkage. 
We only use complete
observations without any missing values, leading to sample sizes between $n
= 599$ and $n = 1,007$ depending on the trait while the number of covariates is always 
$p = 4,410$. 

\begin{figure}
\includegraphics[width = 1\textwidth]{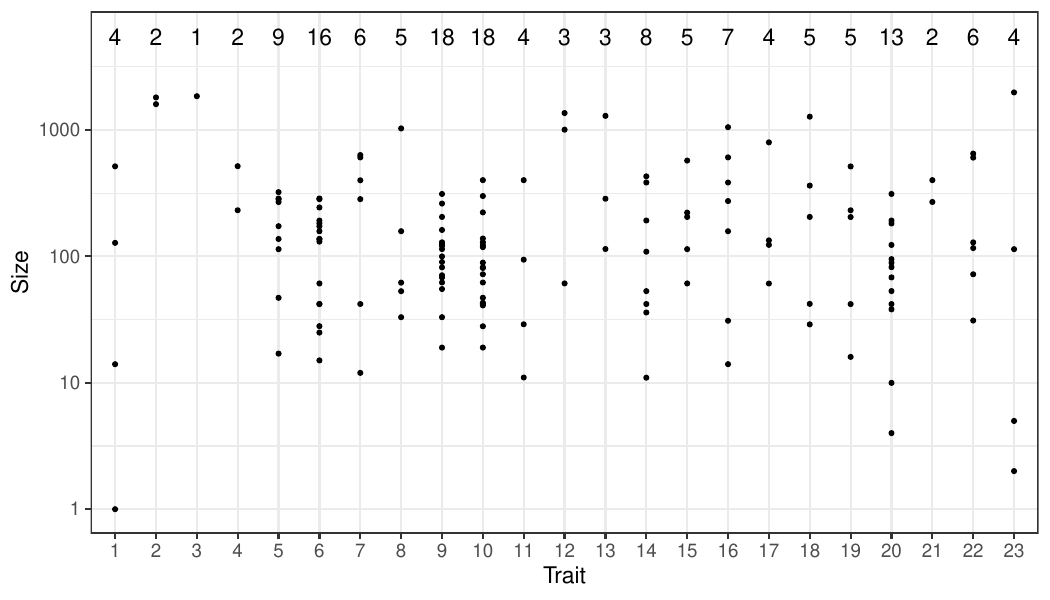}
\caption{Size of significant groups (FWER level $5\%$) by applying
      the hierarchical procedure to each of the first 23 traits of the Yeast
      Colony Growth data set. The number of significant groups is displayed
      on the top. See also Figure \ref{fig.colony supp} in the supplement.} 
          \label{fig.colony}
\end{figure}

{The results across the first 23 traits are given in Figure \ref{fig.colony} and the complete results across all 46 traits are given in Figure \ref{fig.colony supp} in the supplement.}
The hierarchical procedure always finds some
significant groups of SNP covariates. Some of the significant findings
include small groups 
while some of them
are large groups with cardinality bigger than 1, 000. It is plausible that
one cannot find 
many single variables or very small groups but it is reasonable and
convincing to see that the hierarchical method finds a substantial amount
of significant groups. The amount of ``signal'', in terms of significant
findings, varies quite a bit across the 46 traits. 

\vspace{-15pt}

\section*{Acknowledgment} Z. Guo was supported in
    part by the NSF grants DMS-1811857, DMS-2015373 and NIH-1R01GM140463-01; {Z. Guo also acknowledges financial
      support for visiting the Institute of Mathematical Research (FIM) at
      ETH Zurich}.  P. B\"{u}hlmann was supported in part by
    the European Research Council under the Grant Agreement No 786461
    (CausalStats - ERC-2017-ADG);  T.T. Cai was supported in
    part by NSF Grants DMS-1712735 and DMS-2015259 and NIH grants R01-GM129781 and R01-GM123056. Z. Guo is grateful to Dr. Cun-Hui Zhang, Dr. Hongzhe Li, and Dr. Dave Zhao for helpful discussions. 

\section*{Supplementary material}
\label{SM}
Supplementary material is available online includes additional discussion on methods, the technical proofs, additional simulation studies, and an additional real data analysis.


\bibliographystyle{apalike} 
\bibliography{HDRef}

\newpage
\begin{appendices}


\setcounter{page}{1}


\section{Additional Methods and Theories}
\subsection{Inference for $\|\beta_{G}\|_2^2$}
We now consider a commonly used special example by setting $A={\rm I}$ and decompose the error of the plug-in estimator as,
$
\|\widehat{\beta}_{\set}\|_2^2-\|\beta_{\set}\|_2^2=2\langle \widehat{\beta}_{\set},\widehat{\beta}_{\set}-\beta_{\set}\rangle-\|\widehat{\beta}_{\set}-\beta_{\set}\|_2^2.
$
For this special case, the projection direction can actually be identified via the following optimization algorithm, 
\begin{equation*}
\begin{aligned}
\widehat{u}_{\rm I}=\arg\min \;\; u^{\intercal}\widehat{\Sigma} u \quad {\rm s.t.} \;\; \left\|\widehat{\Sigma} u-\begin{pmatrix} \widehat{\beta}_{\set}^{\intercal}& {\bf 0}\end{pmatrix}^{\intercal}\right\|_{\infty} \leq \|\widehat{\beta}_{\set}\|_2 \lambda_{n}.
\end{aligned}
\label{eq: projection I}
\end{equation*}
Note that $\left\|\widehat{\Sigma} u-\begin{pmatrix} \widehat{\beta}_{\set}^{\intercal}& {\bf 0}\end{pmatrix}^{\intercal}\right\|_{\infty}$ can be viewed as $\max_{w\in \mathcal{C}_0}\left\langle w, \widehat{\Sigma}u -\begin{pmatrix} \widehat{\beta}_{\set}^{\intercal}& {\bf 0}\end{pmatrix}^{\intercal}\right\rangle$ where $\mathcal{C}_0=\{e_1,\ldots,e_{p}\}.$ In contrast to $\widehat{u}$ and $\widehat{u}_{\rm A}$, this algorithm for constructing $\widehat{u}_{\rm I}$ is simpler since the constraint set $\mathcal{C}_0$ is smaller than $\mathcal{C}$, that is, we do not need to impose the additional constraint along the direction $\frac{1}{\|\widehat{\beta}_{\set}\|_2}\begin{pmatrix} \widehat{\beta}_{\set}^{\intercal}& {\bf 0} \end{pmatrix}^{\intercal}$. The reason is that $\widehat{\beta}_{\set}$ is close to ${\beta}_{\set}$, which is a sparse vector no matter how large the set $\set$ is.

\subsection{Additional discussion on hierarchical testing}
\label{appendix: hier appl}
Hierarchical testing is a powerful method to go through a sequence of
groups to be tested, from larger groups to smaller ones depending on the
strength of the signal and the amount of correlation among the variables
in and between the groups. As such, it is a multiple testing scheme which
controls the familywise error rate. The details are as follows.

The $p$ covariates are structured into groups of variables in a
hierarchical tree ${\cal T}$ such that at every level of the tree, the groups
build a partition of $\{1,\ldots ,p\}$. 
At a given level, 
the variables in a group have high correlation within groups (and a
tendency for low correlations between groups). The default choice for
constructing such a tree is hierarchical clustering of the
variables \cite[cf.]{hartigan1975clustering}
, typically using 
$1 - \mbox{correlation}^2$ as dissimilarity measure and average linkage.

We assume that the output of hierarchical clustering ${\cal T}$ is 
deterministic, for example when conditioning
on the covariates in the linear model. Hierarchical testing with respect to ${\cal
 T}$ is then a 
sequential multiple testing adjustment procedure as described in the Algorithm 
\ref{al1}. A schematic illustration with a binary hierarchical tree is shown in Figure 2. 

\begin{algorithm}[!h]
\caption{Hierarchical testing procedure} \label{al1}
\begin{tabbing}
   \enspace \textbf{INPUT:} Hierarchical tree ${\cal T}$ with nodes
   corresponding to groups of variables; \\
   \enspace Group testing procedure returning $p$-values $P_G$ for each group of variables $G$, \\
   \enspace e.g. as described in Section \ref{sec: method}; Significance level $\alpha$. \\

   \enspace \textbf{OUTPUT:} Significant groups of variables controlling familywise error rate. \\

\enspace \textbf{REPEAT:}\\
   
   \qquad Go top-down the tree ${\cal T}$ and perform group significance
   testing for groups $G$. \\
   \qquad The raw $p$-value is corrected for multiplicity using \\
   \qquad \qquad \qquad 
   $P_{G;\mbox{adjusted}} = \max_{G' \supseteq G} \tilde{P}_{G'}  \,\,\,\, \mathrm{with } \,\, \tilde{P}_{G} = P_G \cdot p / |G|,$ \\
  
   \qquad where $G'$ is any group in the tree ${\cal T}$. The second line enforces 
   monotonicity of \\ 
   \qquad the adjusted $p$-values. \\

   \qquad For each group $G$ when going top-down in ${\cal T}$: if $P_{G;
     \mbox{adjusted}} \le \alpha$, continue to \\
   \qquad consider the children of
   $G$ for group testing.   \\

   \enspace \textbf{UNTIL:} No more groups are left for testing.\\

\end{tabbing}
\vspace*{-25pt}
\end{algorithm}

\begin{figure}[!htb]

\begin{center}
\begin{tikzpicture}[level distance=0.8cm,
  level 1/.style={sibling distance=5.1cm},
  level 2/.style={sibling distance=3.1cm},
  level 3/.style={sibling distance=2.3cm},
  level 4/.style={sibling distance=1.85cm},
  level 5/.style={sibling distance=1.65cm},
  mycircG/.style={circle, fill=green!60!olive, minimum size=0.5cm},
  mycircR/.style={circle, fill=brown!80!orange, minimum size=0.5cm}
  ] 
\node (z) [mycircG] {   } 
  child {node (a) [mycircG] {   }
    child {node (aa) [mycircG] {   }
    	child {node (aa1) [mycircG] {   }
    		child {node (aa11) [mycircR, label=below:{$\vdots$}] {   }}
    		child {node (ab11) [mycircG] {   }
    			child {node (aa111) [mycircR] {   }}
    			child {node (ab111) [mycircG] {$G_1$}}}
    	}
        child {node (ab1) [mycircR, label=below:{$\vdots$}] {   }}
    }
    child {node (ab) [mycircG] {$G_2$}
    	child {node (aa2) [mycircR, label=below:{$\vdots$}] {   }}
        child {node (ab2) [mycircR, label=below:{$\vdots$}] {   }}
    }
  }
  child {node (c) [mycircG] {   }
    child {node (ca) [mycircR, label=below:{$\vdots$}] {   }}
    child {node (cb) [mycircG] {   }
    	child {node (cb1) [mycircR, label=below:{$\vdots$}] {   }}
    	child {node (cb2) [mycircG] {$G_3$}
        	child {node (cb11) [mycircR, label=below:{$\vdots$}] {   }}
        	child {node (cb12) [mycircR, label=below:{$\vdots$}] {   }}}}
  };
\end{tikzpicture}
 { 
\caption{The hierarchical procedure returns the three groups $G_1$, $G_2$, and $G_3$. 
The green and brown colors highlight significant groups and non-significant groups, 
respectively. 
}}\label{fig1} 
\end{center}
\end{figure}

There are a few interesting properties of hierarchical testing. First, it can be viewed as a hybrid of a sequential procedure and Bonferroni
correction: for every level in the tree, the $p$-value adjustment is a
weighted Bonferroni correction (the standard Bonferroni correction if the
groups have equal size) and across different levels it is a sequential
procedure with no correction but a stopping criterion to not go further
down the tree when no rejection happens. Indeed, the root node needs no
adjustment at all and for each level 
in the tree, the correction depends only on the partitioning on that
level and not on how many tests have been done before. Second, there is no
need to pre-define the level of resolution of the groups. 
The depth is fully data-driven based on the
hierarchical testing procedure. 
Third, the hierarchical
testing method is computationally attractive as no further tests are
considered once a certain group does not exhibit any significance.  \cite{meinshausen2008hierarchical} showed that the procedure controls
the familywise error rate. The hierarchical testing method has
been used for high-dimensional linear models in \cite{manpb16} with a further refinement 
in \cite{manpb16b}
using multi-sample-splitting testing for the groups. The latter is
justified with the strong and questionable assumption that the lasso detects all the relevant variables and in this sense, the
procedure is not fully reliable in terms of error control. See \cite{renaux2020hierarchical} for further details.

\section{Proofs}
\label{sec: proof}
\subsection{Proof of Theorem \ref{thm: Sigma middle}}
\label{sec: thm1 proof}
Throughout the proof, we use the following notations. We use $c$ and $C$ to denote generic positive constants that may vary from place to place. 
For two positive sequences $a_n$ and $b_n$,  $a_n \lesssim b_n$ means $a_n \leq C b_n$ for all $n$,
$a_n \asymp b_n $ if $a_n \lesssim b_n$ and $b_n \lesssim a_n$. 

This proposed estimator $\QWh$ has the following error decomposition, 
\begin{equation*}
\begin{aligned}
\QWh-\QW&=\frac{2}{n}\widehat{u}^{\intercal}X^{\intercal}\epsilon+\beta_{\set}^{\intercal}(\widehat{\Sigma}_{\set,\set}-\Sigma_{\set,\set})\beta_{\set}\\
&+2\left[\widehat{\Sigma}\widehat{u} -(\widehat{\beta}_{\set}^{\intercal}\widehat{\Sigma}_{\set,\set}, {\bf 0})^{\intercal}\right]^{\intercal}(\beta-\widehat{\beta})-(\widehat{\beta}_{\set}-\beta_{\set})^{\intercal}\widehat{\Sigma}_{\set,\set}(\widehat{\beta}_{\set}-\beta_{\set}).
\end{aligned}
\label{eq: Sigma decomposition}
\end{equation*}
We define 
\begin{equation*}
\MW=\frac{2}{n}\widehat{u}^{\intercal}X^{\intercal}\epsilon+\beta_{\set}^{\intercal}(\widehat{\Sigma}_{\set,\set}-\Sigma_{\set,\set})\beta_{\set}
\end{equation*}
and 
\begin{equation*}
\BW=2\left[\widehat{\Sigma}\widehat{u} -(\widehat{\beta}_{\set}^{\intercal}\widehat{\Sigma}_{\set,\set}, {\bf 0})^{\intercal}\right]^{\intercal}(\beta-\widehat{\beta})-(\widehat{\beta}_{\set}-\beta_{\set})^{\intercal}\widehat{\Sigma}_{\set,\set}(\widehat{\beta}_{\set}-\beta_{\set}).
\end{equation*}
Under assumptions 1 and 3, $\epsilon$ is a Gaussian random vector independent of $X$ and $\widehat{u}$ and hence 
\begin{equation}
\frac{2}{n}\widehat{u}^{\intercal}X^{\intercal}\epsilon\mid X,\widehat{u}\sim N\left(0,\frac{4\sigma^2}{n}\widehat{u}^{\intercal}\widehat{\Sigma}\widehat{u}\right).
\label{eq: normality 1}
\end{equation}  
By the central limit theorem and sub-Gaussianity of $X$, we have
\begin{equation}
\beta_{\set}^{\intercal}(\widehat{\Sigma}_{\set,\set}-\Sigma_{\set,\set})\beta_{\set}\cid N\left(0,{\tfrac{1}{n}\E \left({\beta}_{G}^{\intercal} X_{iG} X_{iG}^{\intercal}{\beta}_{G}-{\beta}_{G}^{\intercal}{\Sigma}_{G,G}{\beta}_{G}\right)^2}\right).
\label{eq: normality 2}
\end{equation}
We compute the characteristic function of ${\MW}/{({\rm V}^0_{\Sigma})^{1/2}},$
\begin{equation*}
\begin{aligned}
&\E \exp\left(it {\MW}/{({\rm V}^0_{\Sigma})^{1/2}}\right)\\
&=\E \left(\E \left(\exp(it \MW/({\rm V}^0_{\Sigma})^{1/2})\mid X,\widehat{u}\right)\right)\\
&=\E \left[\E \left(\exp(it \frac{2}{n}\widehat{u}^{\intercal}X^{\intercal}\epsilon/({\rm V}^0_{\Sigma})^{1/2})\mid X,\widehat{u}\right)\cdot \exp(it \beta_{\set}^{\intercal}(\widehat{\Sigma}_{\set,\set}-\Sigma_{\set,\set})\beta_{\set}/({\rm V}^0_{\Sigma})^{1/2})\right]
\end{aligned}
\end{equation*}
By \eqref{eq: normality 1}, we have $$\E \left(\exp(it \frac{2}{n}\widehat{u}^{\intercal}X^{\intercal}\epsilon/({\rm V}^0_{\Sigma})^{1/2})\mid X,\widehat{\beta}\right)=\exp\left(-\frac{t^2}{2}\cdot\frac{\frac{4\sigma^2}{n}\widehat{u}^{\intercal}\widehat{\Sigma}\widehat{u}}{{\rm V}^0_{\Sigma}}\right)$$
Hence 
$$\E \exp\left(it {\MW}/{({\rm V}^0_{\Sigma})^{1/2}}\right)=\E \left[\exp\left(-\frac{t^2}{2}\cdot\frac{\frac{4\sigma^2}{n}\widehat{u}^{\intercal}\widehat{\Sigma}\widehat{u}}{{\rm V}^0_{\Sigma}}\right)\cdot \exp\left(it \frac{\beta_{\set}^{\intercal}(\widehat{\Sigma}_{\set,\set}-\Sigma_{\set,\set})\beta_{\set}}{({\rm V}^0_{\Sigma})^{1/2}}\right)\right]
$$
By \eqref{eq: normality 2} and the condition that $\tfrac{1}{n}\widehat{u}^{\intercal}\widehat{\Sigma}\widehat{u}$ converges in probability to a positive constant $C_1>0$, we have  
$$\frac{\beta_{\set}^{\intercal}(\widehat{\Sigma}_{\set,\set}-\Sigma_{\set,\set})\beta_{\set}}{({\rm V}^0_{\Sigma})^{1/2}} \cid N\left(0, \frac{{\tfrac{1}{n}\E\left({\beta}_{G}^{\intercal} X_{iG} X_{iG}^{\intercal}{\beta}_{G}-{\beta}_{G}^{\intercal}{\Sigma}_{G,G}{\beta}_{G}\right)^2}}{{4\sigma^2}C_1+ {\tfrac{1}{n}\E\left({\beta}_{G}^{\intercal} X_{iG} X_{iG}^{\intercal}{\beta}_{G}-{\beta}_{G}^{\intercal}{\Sigma}_{G,G}{\beta}_{G}\right)^2}}\right)$$ and hence 
$\E \exp\left(it {\MW}/{({\rm V}^0_{\Sigma})^{1/2}}\right)\rightarrow \exp(-\frac{t^2}{2}).$

We control \eqref{eq: remaining} by the following lemma, whose proof is present in Section \ref{sec: lemma proofs}. 
\begin{Lemma} Suppose that Assumptions \ref{assumption1}, \ref{assumption2} and \ref{assumption3} hold, then with probability larger than $1-p^{-c}-g(n)-\exp(-c n^{1/2}),$
\begin{equation}
\left|\left[\widehat{\Sigma}\widehat{u} -(\widehat{\beta}_{\set}^{\intercal}\widehat{\Sigma}_{\set,\set}, {\bf 0})^{\intercal}\right]^{\intercal}(\beta-\widehat{\beta})\right|\lesssim  \|\widehat{\Sigma}_{\set,\set}\widehat{\beta}_{\set}\|_2  \frac{k \log p}{n}
\label{eq: upper 1}
\end{equation}
\begin{equation}
\left|(\widehat{\beta}_{\set}-\beta_{\set})^{\intercal}\widehat{\Sigma}_{\set,\set}(\widehat{\beta}_{\set}-\beta_{\set})\right|\lesssim \|\Sigma_{\set,\set}\|_2 \frac{k \log p}{n}
\label{eq: upper 2}
\end{equation}
\begin{equation}
\left(\frac{1}{n}\widehat{u}^{\intercal}X^{\intercal}X\widehat{u}\right)^{1/2}\asymp  \|\widehat{\Sigma}_{\set,\set}\widehat{\beta}_{\set}\|_2 
\label{eq: variance}
\end{equation}
\label{lem: key lemma 1}
\end{Lemma}
Then the control of the reminder term \eqref{eq: remaining} follows from \eqref{eq: upper 1} and \eqref{eq: upper 2}. By the expression of $V_{\Sigma}$ in \eqref{eq: implication}, we then establish $\left|\BW\right|\leq \eta z_{1-\alpha/2} (\VW)^{1/2}$ with a high probability by applying \eqref{eq: remaining}, \eqref{eq: variance} and the condition $k\leq c {n^{1/2}}/{\log p}$ for some positive constant $c>0.$ As a consequence, we establish \eqref{eq: implication}.

\subsection{Proofs of Theorem \ref{thm: known middle}}
The proof of Theorem \ref{thm: known middle} is similar to that of Theorem \ref{thm: Sigma middle}. We start with the decomposition
\begin{equation*}
\begin{aligned}
\QAh-\QA=\frac{2}{n}\widehat{u}_{A}^{\intercal}X^{\intercal}\epsilon
+2\left[\widehat{\Sigma}\widehat{u}_{A} -\begin{pmatrix} \widehat{\beta}_{\set}^{\intercal}A& {\bf 0}\end{pmatrix}^{\intercal}\right]^{\intercal}(\beta-\widehat{\beta})-(\widehat{\beta}_{\set}-\beta_{\set})^{\intercal}A(\widehat{\beta}_{\set}-\beta_{\set}).
\end{aligned}
\label{eq: known decomposition}
\end{equation*}
and define
\begin{equation*}
\MA=\frac{2}{n}\widehat{u}_{A}^{\intercal}X^{\intercal}\epsilon
\end{equation*}
and 
\begin{equation*}
\BA=2\left[\widehat{\Sigma}\widehat{u}_{A} -\begin{pmatrix} \widehat{\beta}_{\set}^{\intercal}A& {\bf 0}\end{pmatrix}^{\intercal}\right]^{\intercal}(\beta-\widehat{\beta})-(\widehat{\beta}_{\set}-\beta_{\set})^{\intercal}A(\widehat{\beta}_{\set}-\beta_{\set})
\end{equation*}
We can establish a similar Lemma as Lemma \ref{lem: key lemma 1} and present the corresponding proof in Section \ref{sec: lemma proofs} of the supplementary materials.

\begin{Lemma} Suppose that Assumptions \ref{assumption1}, \ref{assumption2} and \ref{assumption3} hold, then with probability larger than $1-p^{-c}-g(n,p),$ for some positive constant $c>0$,
\begin{equation}
\left|\left[\widehat{\Sigma}\widehat{u}_{A} -\begin{pmatrix} \widehat{\beta}_{\set}^{\intercal}A& {\bf 0}\end{pmatrix}^{\intercal}\right]^{\intercal}(\beta-\widehat{\beta})\right|\lesssim  \|A\widehat{\beta}_{\set}\|_2  \frac{k \log p}{n}
\label{eq: upper 1 known}
\end{equation}
\begin{equation}
\left|(\widehat{\beta}_{\set}-\beta_{\set})^{\intercal}A(\widehat{\beta}_{\set}-\beta_{\set})\right|\lesssim \|A\|_2 \frac{k \log p}{n}
\label{eq: upper 2 known}
\end{equation}
\begin{equation}
\left(\frac{1}{n}\widehat{u}_{A}^{\intercal}X^{\intercal}X\widehat{u}_{A}\right)^{1/2}\asymp  \|A\widehat{\beta}_{\set}\|_2 
\label{eq: variance known}
\end{equation}
\label{lem: key lemma 2}
\end{Lemma}
Then we establish the asymptotic normality of $M_{A}$ by the fact that $\epsilon_i$ are normal random variables. The high probability bound in \eqref{eq: remaining known} follows from \eqref{eq: upper 1 known} and \eqref{eq: upper 2 known}. Then \eqref{eq: implication known} follows from the fact that $\left|\BA\right|\leq \eta z_{1-\alpha/2} ({\VA})^{1/2}$, where this inequality follows from \eqref{eq: variance known} and the condition $k\leq c {n^{1/2}}/{\log p}$ for some positive constant $c>0$.
 
\subsection{Proofs of Lemmas \ref{lem: key lemma 1} and \ref{lem: key lemma 2}}
\label{sec: lemma proofs}
The proof of \eqref{eq: upper 1} follows from 
$$\left|\left[\widehat{\Sigma}\widehat{u} -(\widehat{\beta}_{\set}^{\intercal}\widehat{\Sigma}_{\set,\set}, {\bf 0})^{\intercal}\right]^{\intercal}(\beta-\widehat{\beta})\right|
\leq \|\widehat{\Sigma}\widehat{u} -(\widehat{\beta}_{\set}^{\intercal}\widehat{\Sigma}_{\set,\set}, {\bf 0})^{\intercal}\|_{\infty}\|\beta-\widehat{\beta}\|_1$$
together with the constraint \eqref{eq: projection Cov} and Assumption \ref{assumption2}.
The proof of \eqref{eq: upper 1 known} follows from
$$\left|\left[\widehat{\Sigma}\widehat{u}_{A} -\begin{pmatrix} \widehat{\beta}_{\set}^{\intercal}A& {\bf 0}\end{pmatrix}^{\intercal}\right]^{\intercal}(\beta-\widehat{\beta})\right|\leq \|\widehat{\Sigma}\widehat{u}_{A} -\begin{pmatrix} \widehat{\beta}_{\set}^{\intercal}A& {\bf 0}\end{pmatrix}^{\intercal}\|_{\infty} \|\beta-\widehat{\beta}\|_1$$
together with the constraint for constructing $\widehat{u}_{A}$ and Assumption \ref{assumption2}.

The proof of \eqref{eq: upper 2 known} follows from $\left|(\widehat{\beta}_{\set}-\beta_{\set})^{\intercal}A(\widehat{\beta}_{\set}-\beta_{\set})\right|\leq \|A\|_2\|\widehat{\beta}_{\set}-\beta_{\set}\|_2^2$ and Assumption \ref{assumption2}. The proof of \eqref{eq: upper 2} follows from Lemma 11 of \cite{cai2018semi}, specifically, the definition of event $\set_{6}(\widehat{\beta}_{\set}-\beta_{\set},\widehat{\beta}_{\set}-\beta_{\set},n^{1/2})$ and hence with probability larger than $1-\exp(-n^{1/2}),$
$$\left|(\widehat{\beta}_{\set}-\beta_{\set})^{\intercal}\widehat{\Sigma}_{\set,\set}(\widehat{\beta}_{\set}-\beta_{\set})\right|\lesssim \left|(\widehat{\beta}_{\set}-\beta_{\set})^{\intercal}{\Sigma}_{\set,\set}(\widehat{\beta}_{\set}-\beta_{\set})\right|\leq \|\Sigma_{\set,\set}\|_2\|\widehat{\beta}_{\set}-\beta_{\set}\|_2^2.$$

Under the independence assumption imposed Assumption \ref{assumption3}, 
we apply Lemma 1 of \cite{cai2019individualized} by taking $x_{\rm new}=(\widehat{\beta}_{\set}^{\intercal}\widehat{\Sigma}_{\set,\set}, {\bf 0})^{\intercal}$ and consider the one sample setting and then we establish \eqref{eq: variance}. Similarly, we can also establish \eqref{eq: variance known}.

\subsection*{Proofs of Corollaries \ref{cor: 1}, \ref{cor: 2} and \ref{cor: local heritability}}
\label{sec: cor proofs}
Define $d_n(\tau)=\frac{\sqrt{{\VWh}}}{\sqrt{{\VW(\tau)}}}-1.$
By Assumption \ref{assumption2} and $k \log p \lesssim n^{1/2}$, we have 
\begin{equation}
\frac{\sqrt{{\VWh}}}{\sqrt{{\VW(\tau)}}}\cip 1.
\label{eq: var consistency}
\end{equation}
The above variance consistency result is implied by Lemma 4 of \cite{cai2018semi}. In particularly, we apply Lemma 4 of \cite{cai2018semi} with taking $\rho$ therein as 1 and the following equivalent expression,
$$\E \left({\beta}_{G}^{\intercal} X_{iG} X_{iG}^{\intercal}{\beta}_{G}-{\beta}_{G}^{\intercal}{\Sigma}_{G,G}{\beta}_{G}\right)^2=\E \left(\begin{pmatrix}{\beta}_{G}^{\intercal} &{\bf 0}^{\intercal}\end{pmatrix} X_{i,\cdot} X_{i,\cdot}^{\intercal}\begin{pmatrix}{\beta}_{G}^{\intercal} &{\bf 0}^{\intercal}\end{pmatrix}^{\intercal}-\begin{pmatrix}{\beta}_{G}^{\intercal} &{\bf 0}^{\intercal}\end{pmatrix}{\Sigma}\begin{pmatrix}{\beta}_{G}^{\intercal} &{\bf 0}^{\intercal}\end{pmatrix}^{\intercal}\right)^2.$$
 
Note that 
$$
\mathbf{pr}_{\theta}\left(\phi_{\Sigma}(\tau)=1\right)=\mathbf{pr}_{\theta}\left(\QWh\geq z_{1-\alpha}\sqrt{{\VWh}}\right)=\mathbf{pr}_{\theta}\left(\QW+\MW+\BW\geq z_{1-\alpha}\sqrt{{\VWh}}\right).$$
Together with the definition of $d_{n}(\tau)$, we can further control the above probability by 
\begin{equation}
\begin{aligned}
&\mathbf{pr}_{\theta}\left(\MW\geq (1+d_n(\tau))z_{1-\alpha}\sqrt{{\VW}}+\eta z_{1-\alpha} (1+d_n(\tau))\sqrt{{\VW}}-{\BW}-\QW\right)\\
&=\mathbf{pr}_{\theta}\left(\frac{\MW}{\sqrt{{\VW}}}\geq (1+d_n(\tau))z_{1-\alpha}+\eta z_{1-\alpha} (1+d_n(\tau))-\frac{\BW}{\sqrt{{\VW}}}-\frac{\QW}{\sqrt{{\VW}}}\right)\\
\end{aligned}
\label{eq: general power}
\end{equation}
Then we control the type I error in Corollary \ref{cor: 1}, following from the limiting distribution established in Theorem \ref{thm: Sigma middle} and the fact that $\eta z_{1-\alpha} (1+d_n(\tau))-\frac{\BW}{\sqrt{{\VW}}}$ is asymptotically positive under the condition $k \lesssim {n^{1/2}}/{\log p}$. We can also establish the lower bound for the asymptotic power in Corollary \ref{cor: 2} by \eqref{eq: general power}, the definition 
$\delta(t)=((1+2\eta)z_{1-\alpha}+t)(\V_{\Sigma})^{1/2}$ and the fact that $\eta z_{1-\alpha} (1+d_n(\tau))-\frac{\BW}{\sqrt{{\VW}}}$ is asymptotically positive under the condition $k \lesssim {n^{1/2}}/{\log p}.$
We can use the same argument to control the type I error and the asymptotic power of $\phi_{A}(\tau)$.

The proof of Corollary \ref{cor: local heritability} follows from \eqref{eq: implication} and \eqref{eq: implication known}, Assumption \ref{assumption2} that $\widehat{\sigma}^2$ is a consistent estimator of $\sigma^2$ and \eqref{eq: var consistency}.

\section{Additional Numerical Results}
\label{sec: add sim}

\subsection{Additional Simulations for Dense Alternative Setting}
\label{sec: add dense}

We take the same simulation setting for dense alternative as in Section \ref{sec: sim dense} in the main paper and vary the signal strength parameter $\delta$ over $\{0, 0.04,0.06, 0.08, 0.2,0.4\}$ and the sample size $n$ over $\{250,350,500,650\}$.  We examine the finite sample performance of the proposed method over different values of $\tau \in \{0,0.1,0.2,\cdots,1.4,1.5\}.$  The main observations are similar to those reported in Section \ref{sec: sim dense} in the main paper. We shall focus more on how $\tau$ will affect the proposed inference methods.

\begin{figure}
    \centering
    \subfigure[ERR of $\phi_{\rm I}(\tau)$ defined in \eqref{eq: test general}]{\includegraphics[width = 1\textwidth]{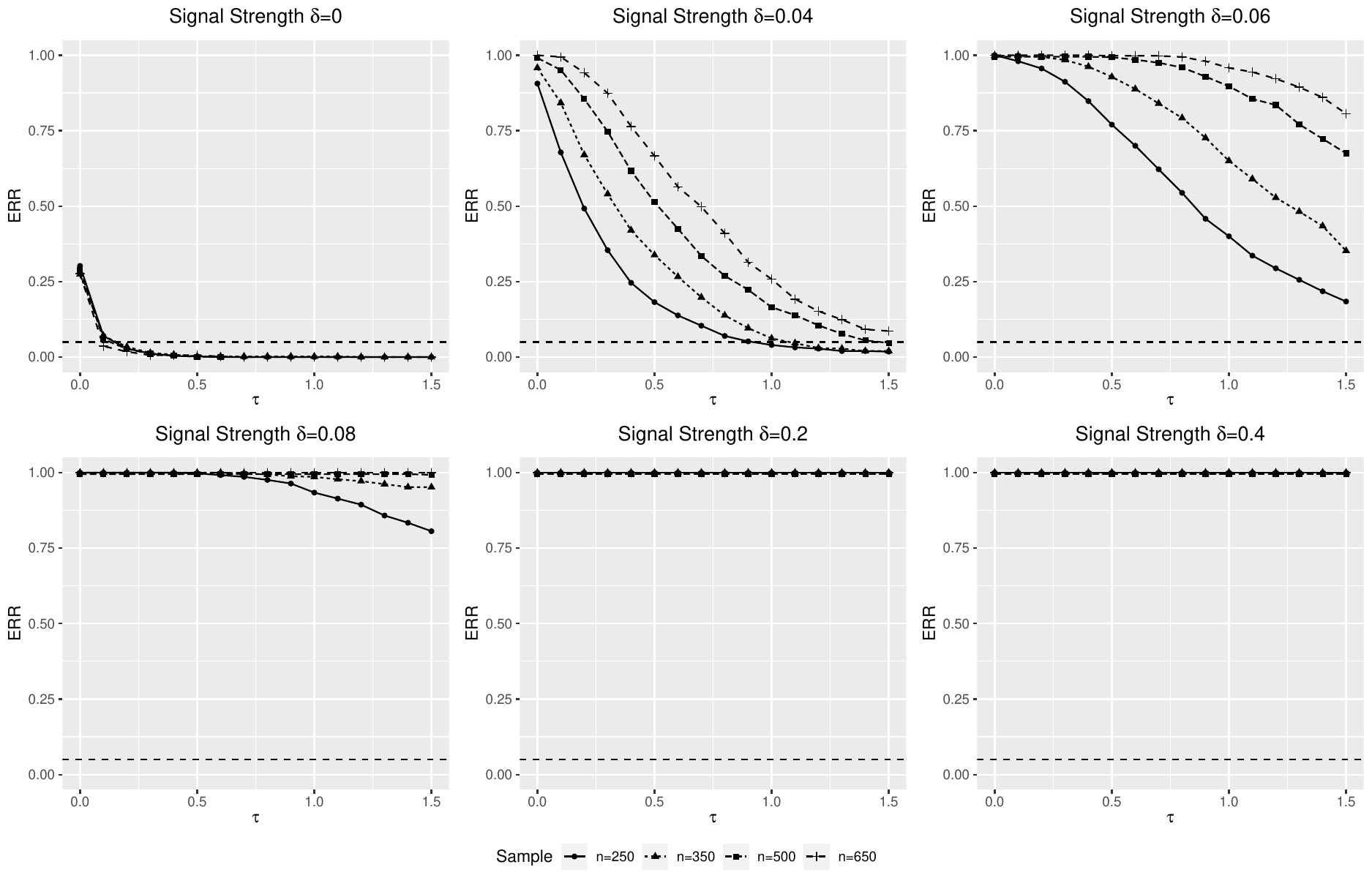}}%
    \qquad
	\subfigure[ERR of $\phi_{\Sigma}(\tau)$ defined in \eqref{eq: test Cov}]{\includegraphics[width = 1\textwidth]{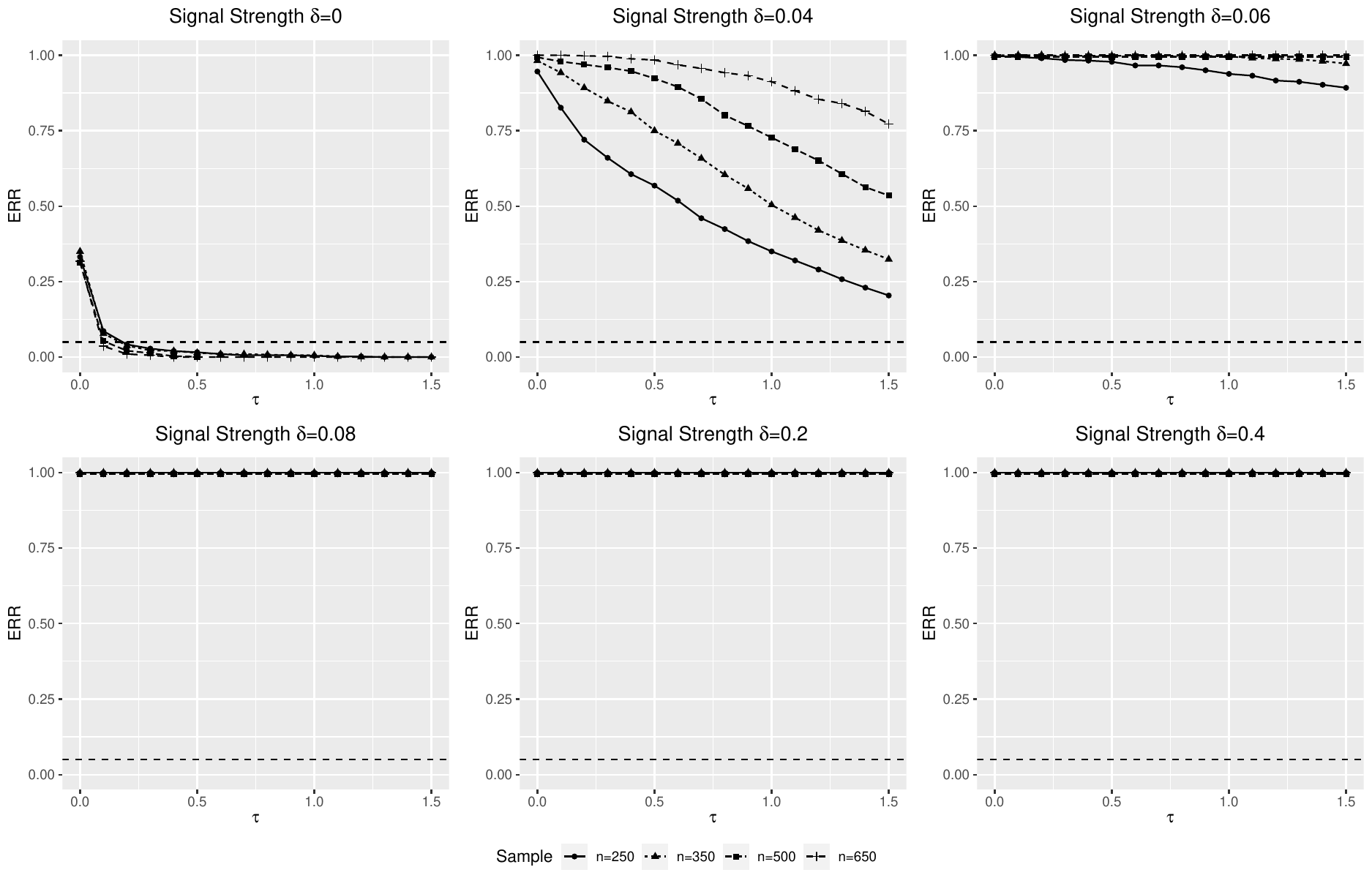}}%
    \caption{Dense alternative setting: the dependence of Empirical Rejection Rate (ERR)  on $\tau$.}%
    \label{fig: dense-det}%
\end{figure}

We report the empirical rejection rate for $\phi_{\rm I}(\tau)$ and $\phi_{\Sigma}(\tau)$ in Figure \ref{fig: dense-det}. The test $\phi_{\Sigma}$ is in general more powerful than $\phi_{\rm I}.$ It is observed that for the null setting with $\delta=0$, the testing procedure with $\tau=0$ does not guarantee the type I error while the type I error is controlled as long as $\tau$ reaches $0.1$. We have seen that for a small $\delta \in \{0.04,0.06, 0.08\}$, the choice of $\tau$ has a relatively strong effect on the power of $\phi_{\rm I}$; the larger the $\tau$ value, the less powerful the test is. When the signal strength is large enough (that is, $\delta=0.2, 0.4$), the effect of $\tau$ on the powers of the tests $\phi_{\Sigma}$ and $\phi_{\rm I}$ are marginal.
 
\begin{figure}%
    \centering
    \subfigure[Empirical Coverage of ${\rm CI}_{\rm I}(\tau)$ defined in \eqref{eq: CI general}]{\includegraphics[width = 1\textwidth]{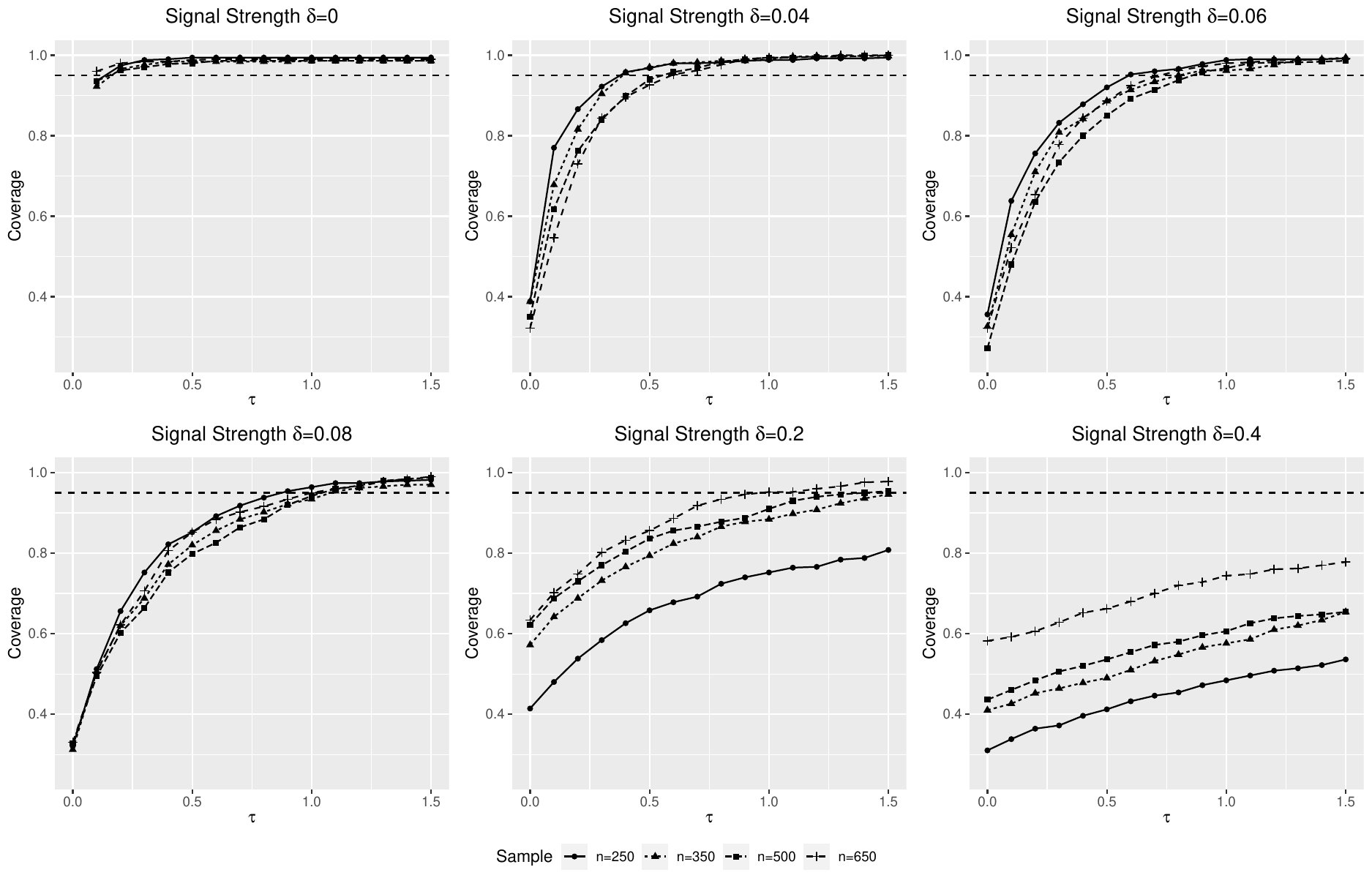}}%
    \qquad
	\subfigure[Empirical Coverage of ${\rm CI}_{\Sigma}(\tau)$ defined in \eqref{eq: CI Cov}.]{\includegraphics[width = 1\textwidth]{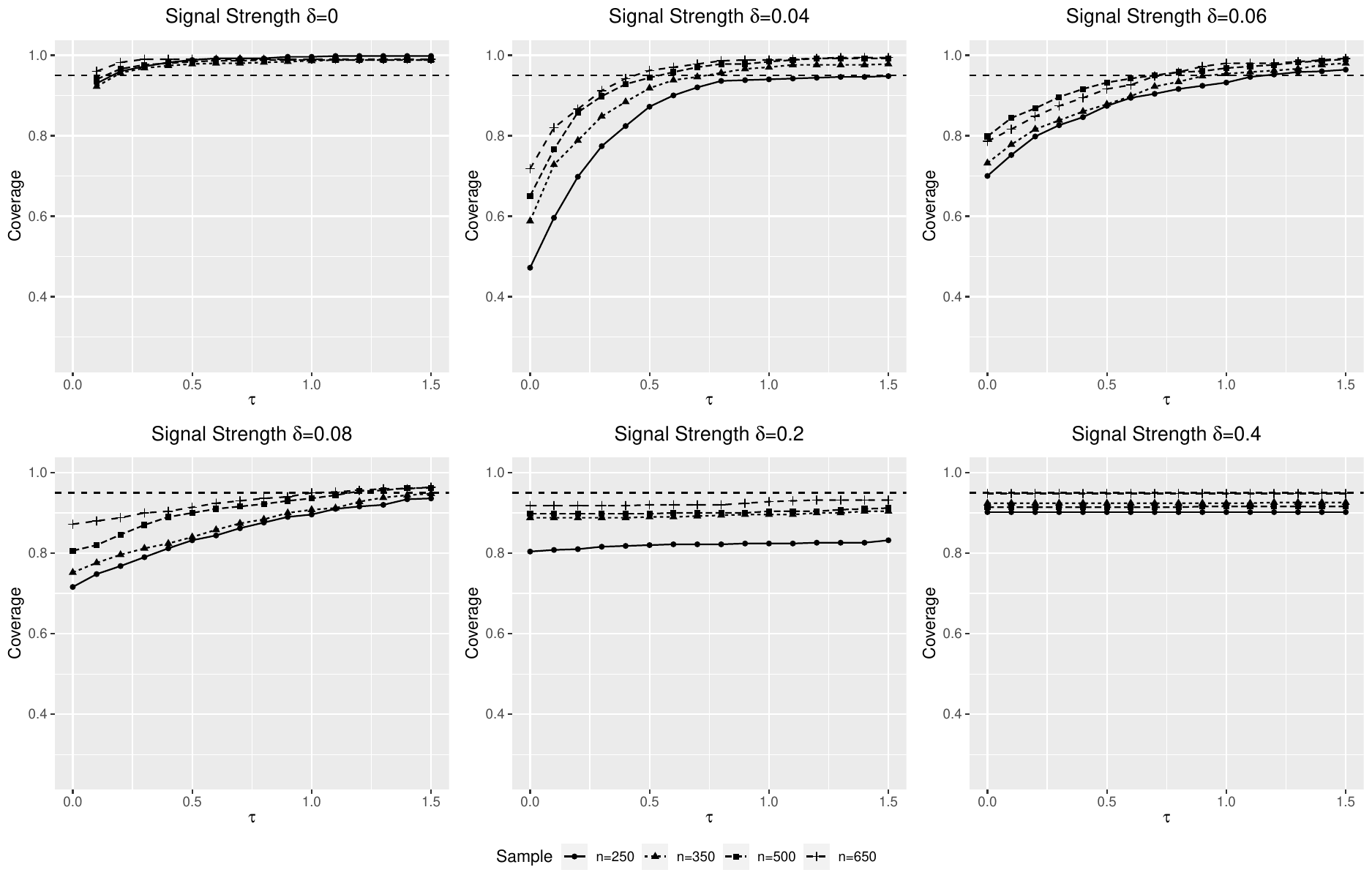}}%
    \caption{Dense alternative setting: the dependence of Empirical Coverage on $\tau$.}%
    \label{fig: dense-cov}%
\end{figure}

We report the coverage properties of the constructed confidence intervals ${\rm CI}_{\rm I}(\tau)$ and ${\rm CI}_{\Sigma}(\tau)$ in Figure \ref{fig: dense-cov}. It is observed that, for $\tau=1$, the empirical coverage of the constructed confidence intervals reaches the $95\%$ level, which is the dashed line plotted in each plot. For most cases, $\tau=0.5$ leads to reasonable coverage properties though the empirical coverage properties do not always reach $95\%$. The empirical coverage of ${\rm CI}_{\Sigma}(\tau)$ is general better than that of ${\rm CI}_{\rm I}(\tau).$ This reflects that the inference problem for $\|\beta_{G}\|_2^2$ might be harder due to inverting $\Sigma_{G,G}$ in the construction of the projection direction. We point out that we only report the empirical coverage above $0.25$. Specifically, for $\delta=0$, the empirical coverage of ${\rm CI}_{\rm I}(\tau)$ and ${\rm CI}_{\Sigma}(\tau)$ with $\tau=0$ are not plotted as their values are below $0.25.$

We shall highlight two interesting observations with further explanations. Firstly, for ${\rm CI}_{\Sigma}(\tau)$, when $\delta$ is relatively large, say $0.2$ or $0.4$, the choice of $\tau$ does not affect the empirical coverage much. This matches with the established theoretical results in Theorem \ref{thm: Sigma middle} that, when $\beta$ has a relatively large norm value, the super-efficiency phenomenon disappears. That is, even for $\tau=0$, the variance of $\QWh$ is of order $1/\sqrt{n}$ and dominates the bias in the setting of a sufficiently sparse $\beta.$ 

Secondly, we observe that the coverage property of ${\rm CI}_{\rm I}(\tau)$ for $\delta=0.4$ is bad even when $n=650$. This under-coverage happens due to the large sparsity of this simulation setting. Note that there are $21$ non-zero coordinates in the simulation setting and for small $\delta$, its effective sparsity level (e.g. the capped $\ell_1$ sparsity) can be smaller than $21$. However, for the relatively strong signal $\delta=0.4$, the effective sparsity is large and violates the key sparsity assumption $k\leq c\sqrt{n}/\log p$ in  Theorem \ref{thm: known middle}. As a further remark, for $\delta=0.4$, the center of the confidence interval ${\rm CI}_{\rm I}(\tau)$ is still close to $\|\beta_{G}\|_2^2$ but the uncertainty quantification is not accurate enough since we only quantify the uncertainty of the asymptotic normal component. In Section \ref{sec: sparser setting}, we consider a similar simulation setting with a reduced sparsity level and observe that the constructed confidence intervals achieve the desired coverage level for $\delta=0.4$.

\subsection{Additional Simulations for High Correlation Setting}

We take the same simulation setting as the high correlation setting in Section \ref{sec: sim high-cor} in the main paper. We vary the signal strength parameter $\delta$ over $\{0, 0.1, 0.2, 0.3, 0.4, 0.5\}$ and the sample size $n$ over $\{250,350,500,650\}$. We have examined the finite sample performance of the proposed methods over different values of $\tau \in \{0,0.1,0.2,\cdots,1.4,1.5\}.$ The main observations are similar to those reported in Section \ref{sec: sim high-cor} in the main paper. 

\begin{figure}%
    \centering
    \subfigure[ERR of $\phi_{\rm I}(\tau)$ defined in \eqref{eq: test general}]{\includegraphics[width = 1\textwidth]{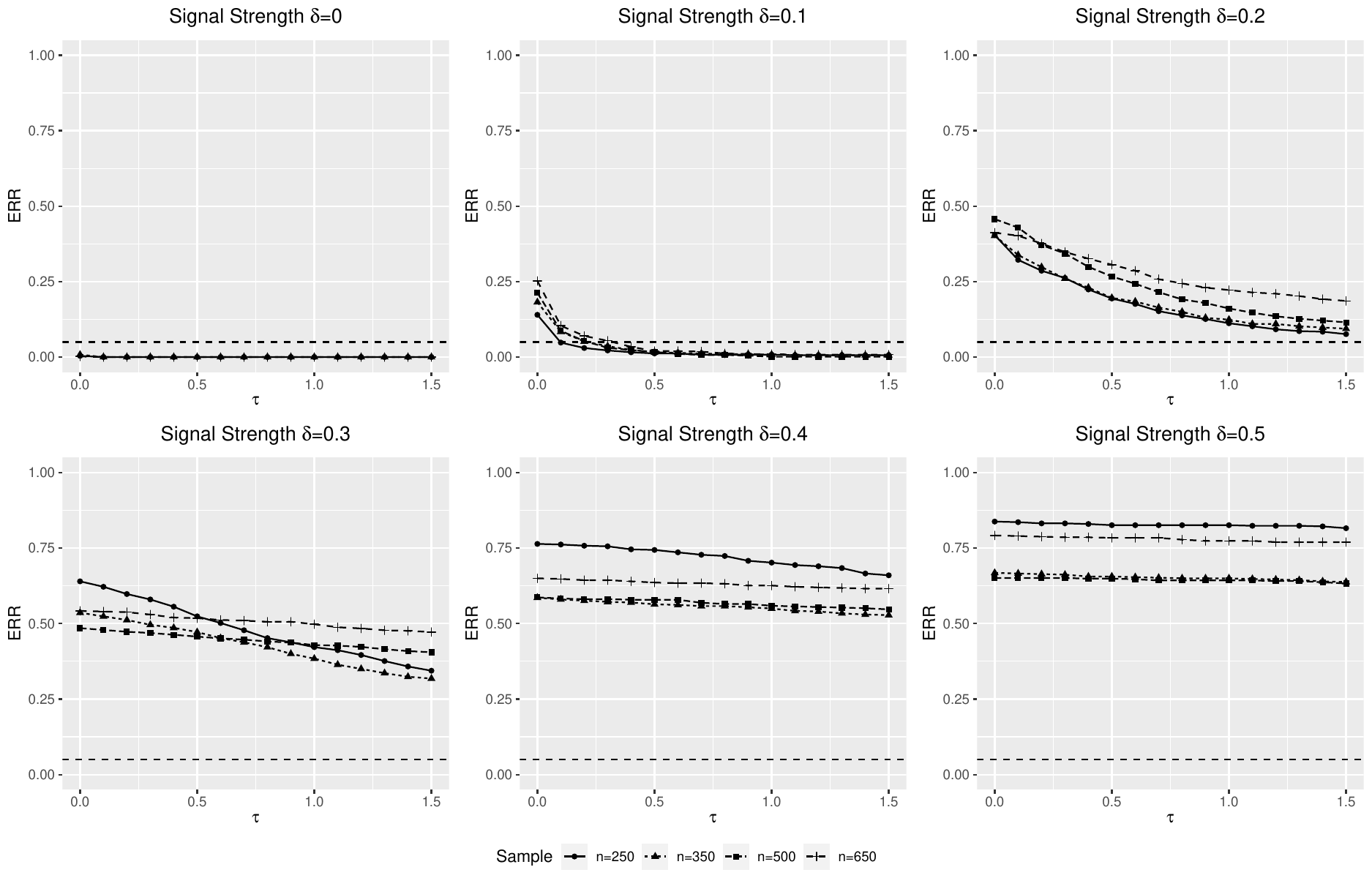}}%
    \qquad
	\subfigure[ERR of $\phi_{\Sigma}(\tau)$ defined in \eqref{eq: test Cov}]{\includegraphics[width = 1\textwidth]{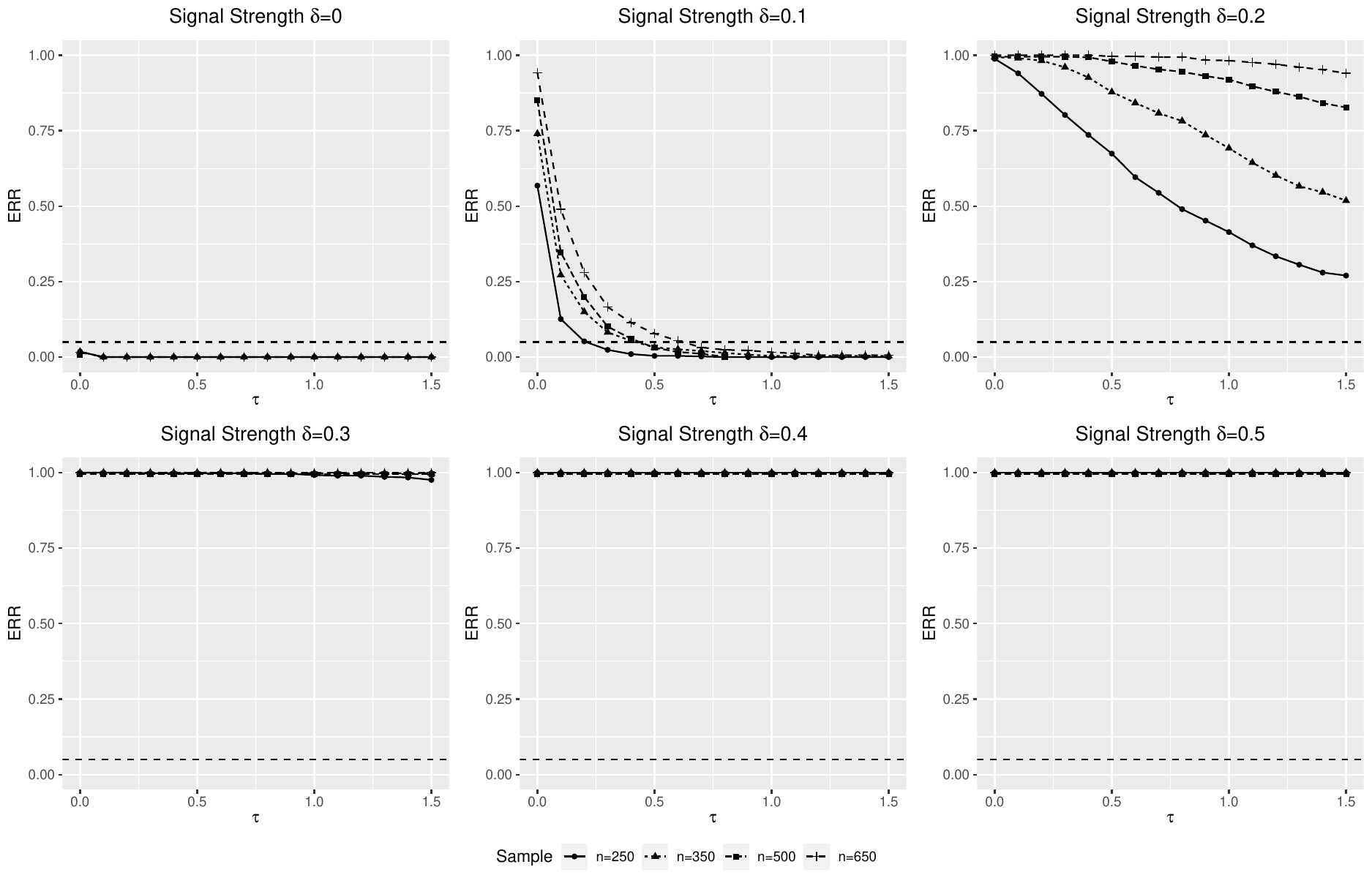}}%
    \caption{High correlation setting: the dependence of Empirical Rejection Rate (ERR)  on $\tau$.}%
    \label{fig: HC-det}%
\end{figure}

\begin{figure}%
    \centering
    \subfigure[Empirical Coverage of ${\rm CI}_{\rm I}(\tau)$ defined in \eqref{eq: CI general}]{\includegraphics[width = 1\textwidth]{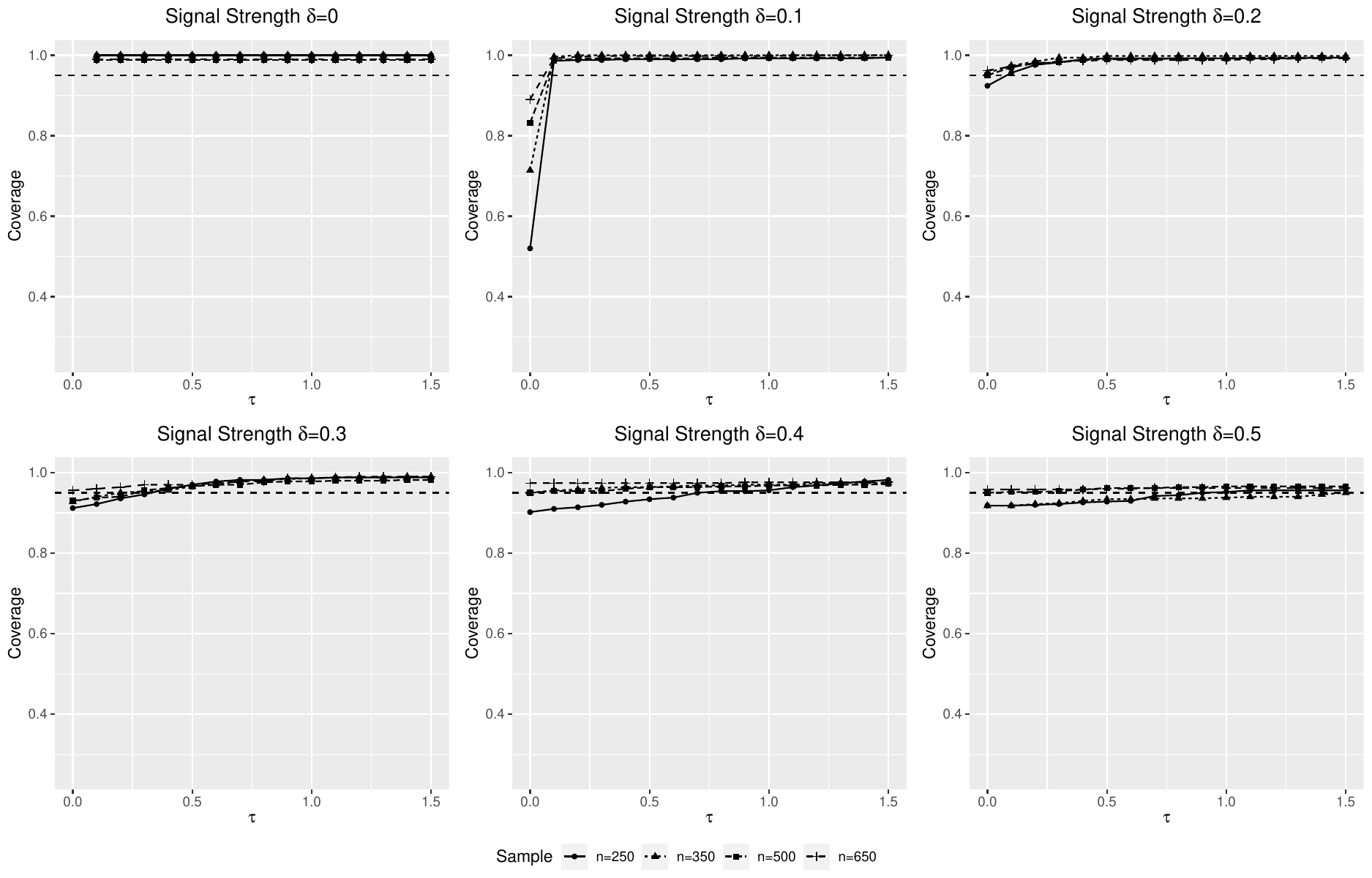}}%
    \qquad
	\subfigure[Empirical Coverage of ${\rm CI}_{\Sigma}(\tau)$ defined in \eqref{eq: CI Cov}.]{\includegraphics[width = 1\textwidth]{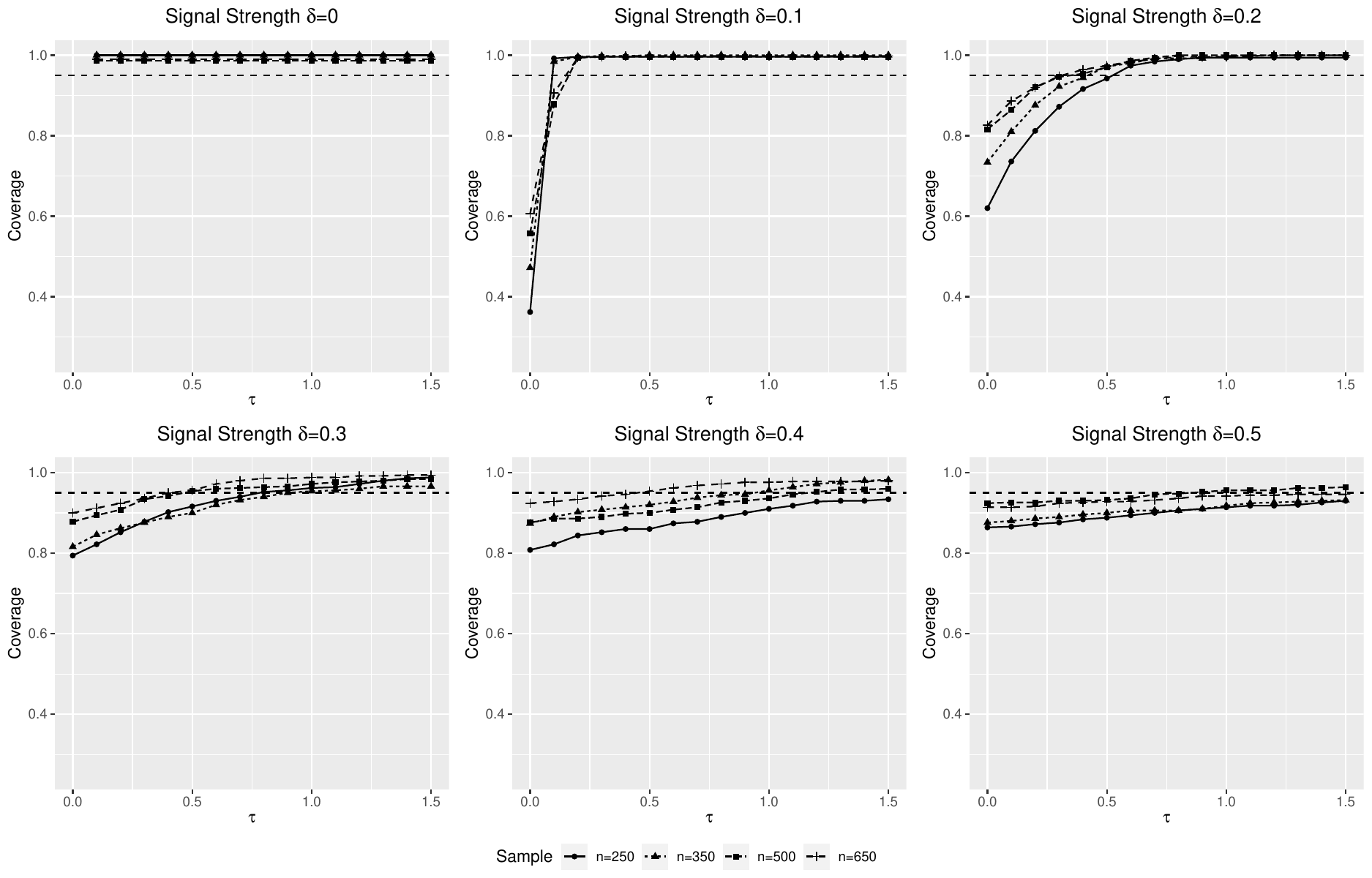}}%
    \caption{High correlation setting: the dependence of Empirical Coverage on $\tau$.}%
    \label{fig: HC-cov}%
\end{figure}

Regarding the effect $\tau$, the observation is similar to the dense alternative setting reported in Section \ref{sec: add dense}. $\tau=0.5$ or $\tau=1$ leads to reliable testing and coverage properties and when $\delta$ is above $0.3$, the effect of $\tau$ is marginal. We report the empirical rejection rate for $\phi_{\rm I}(\tau)$ and $\phi_{\Sigma}(\tau)$ in Figure \ref{fig: HC-det} and the coverage properties of the constructed confidence intervals ${\rm CI}_{\rm I}(\tau)$ and ${\rm CI}_{\Sigma}(\tau)$ in Figure \ref{fig: HC-cov}.  The main observations are similar to those in Section \ref{sec: add dense}. We observe that the test $\phi_{\Sigma}$ is in general more powerful than $\phi_{\rm I}$ and the empirical coverage of both ${\rm CI}_{\rm I}(\tau)$ and ${\rm CI}_{\Sigma}(\tau)$ reaches the desired level even for the high correlation setting.

\subsection{Dependence on sparsity}
\label{sec: sparser setting}
We take the same simulation as the dense alternative simulation in Section \ref{sec: sim dense}, except for generating a sparser regression vector $\beta$: $\beta_j=\sqrt{7}\cdot \delta$ for $30\leq j\leq 32$  and $\beta_j=0$ otherwise.  We consider the same group significance test as in Section \ref{sec: sim dense},
$
H_{0,G}: \beta_{i}=0 \; \text{for} \; i\in G, $ with  $G=\{30,31,\cdots,200\}. 
$
The rescaling parameter $\sqrt{7}$ in generating $\beta$ guarantees the same values of $\|\beta_{G}\|_2^2$ and $\beta_{G}^{\intercal}\Sigma_{G,G}\beta_{G}$ as in the dense alternative setting in Section \ref{sec: sim dense}. We vary the signal strength parameter $\delta$ over $\{0, 0.04,0.06, 0.08, 0.2,0.4\}$ and the sample size $n$ over $\{250,350,500,650\}$. We have examined the finite sample performance of the proposed method over different values of $\tau \in \{0,0.1,0.2,\cdots,1.4,1.5\}.$ 

\begin{figure}%
    \centering
    \subfigure[ERR of $\phi_{\rm I}(\tau)$ defined in \eqref{eq: test general}]{\includegraphics[width = 1\textwidth]{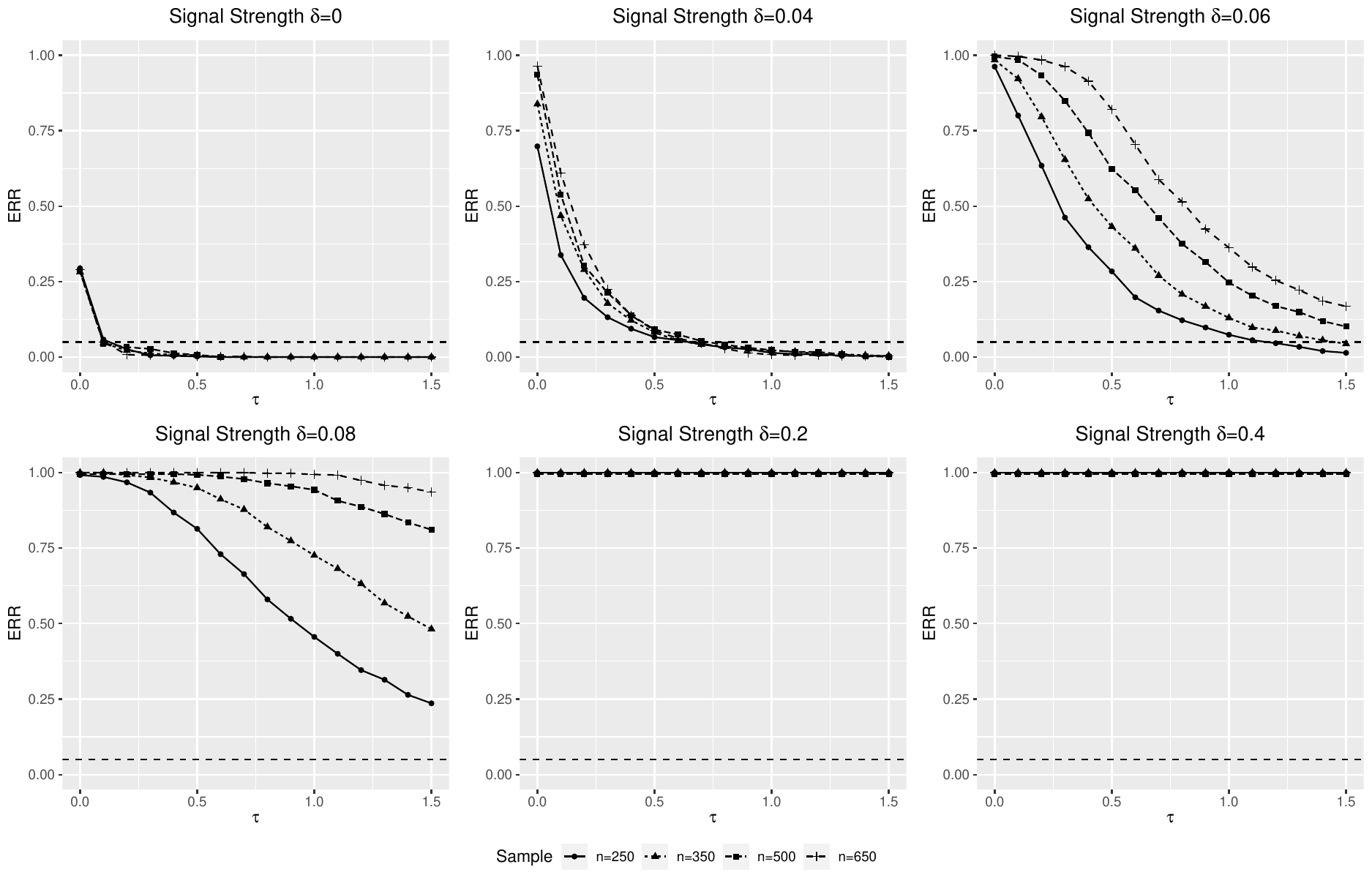}}%
    \qquad
	\subfigure[ERR of $\phi_{\Sigma}(\tau)$ defined in \eqref{eq: test Cov}]{\includegraphics[width = 1\textwidth]{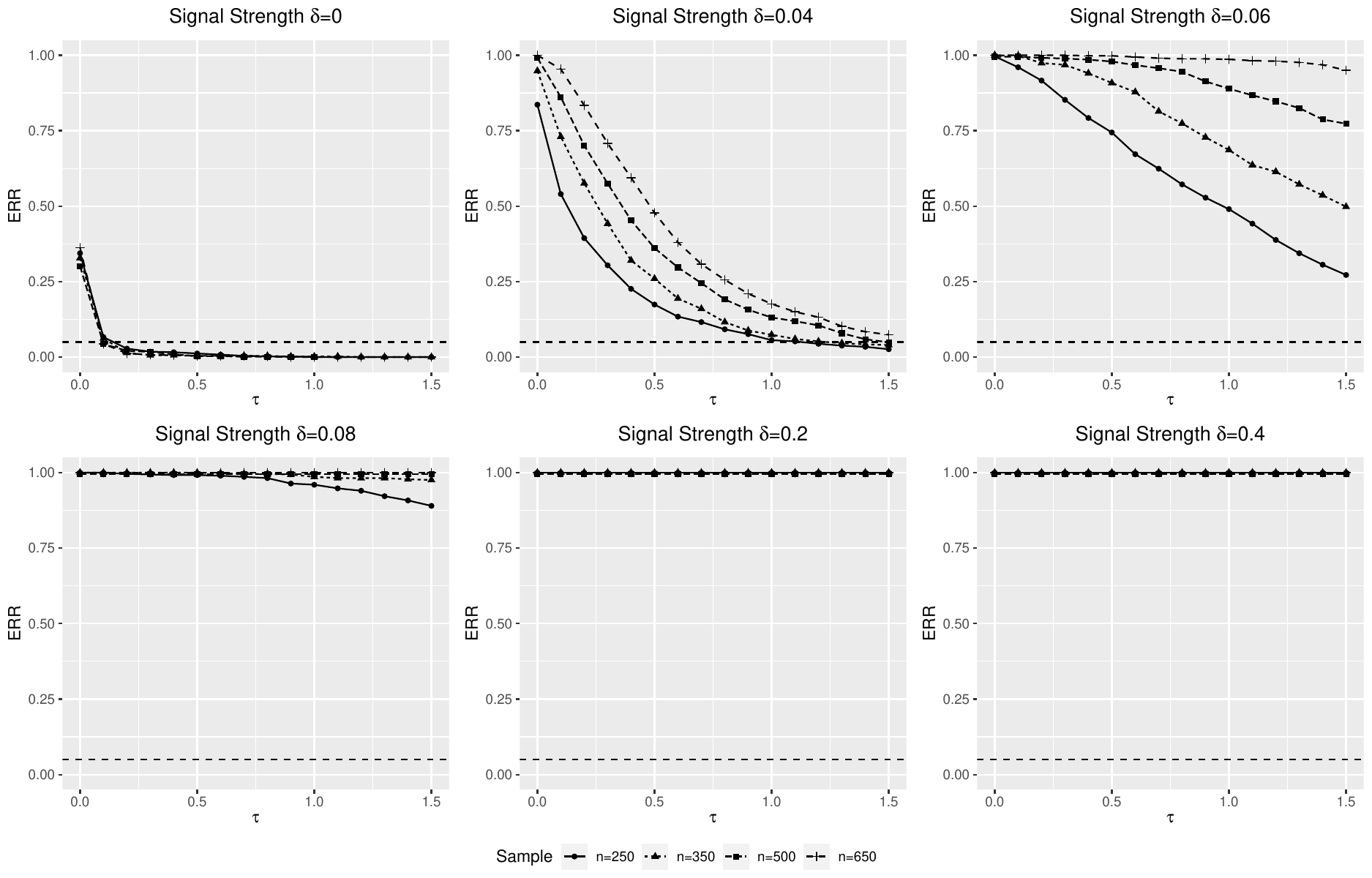}}%
    \caption{Dense alternative setting with a sparser signal: the dependence of Empirical Rejection Rate (ERR)  on $\tau$.}%
    \label{fig: dense-det-sparser}%
\end{figure}

\begin{figure}%
    \centering
    \subfigure[Empirical Coverage of ${\rm CI}_{\rm I}(\tau)$ defined in \eqref{eq: CI general}]{\includegraphics[width = 1\textwidth]{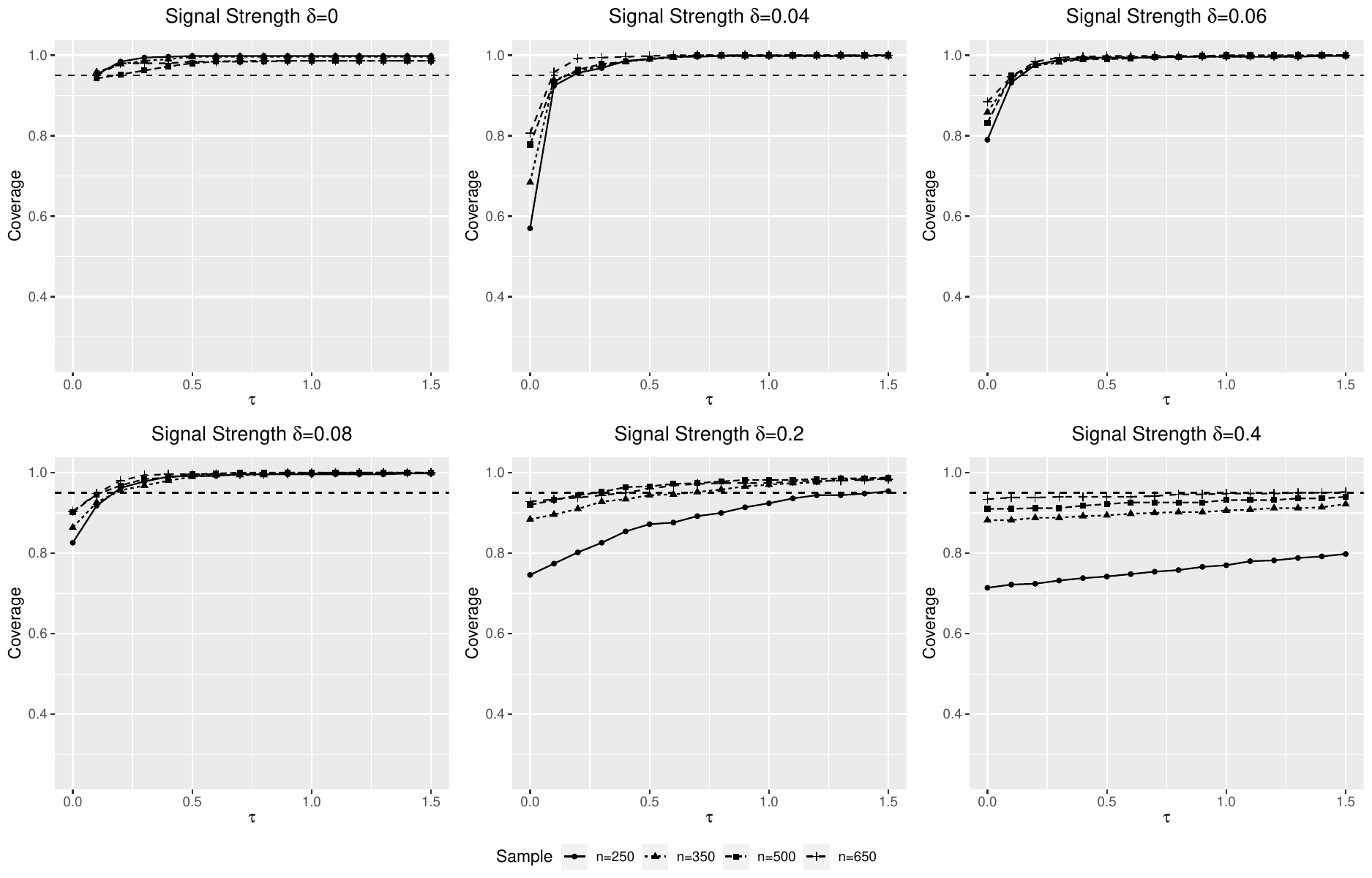}}%
    \qquad
	\subfigure[Empirical Coverage of ${\rm CI}_{\Sigma}(\tau)$ defined in \eqref{eq: CI Cov}.]{\includegraphics[width = 1\textwidth]{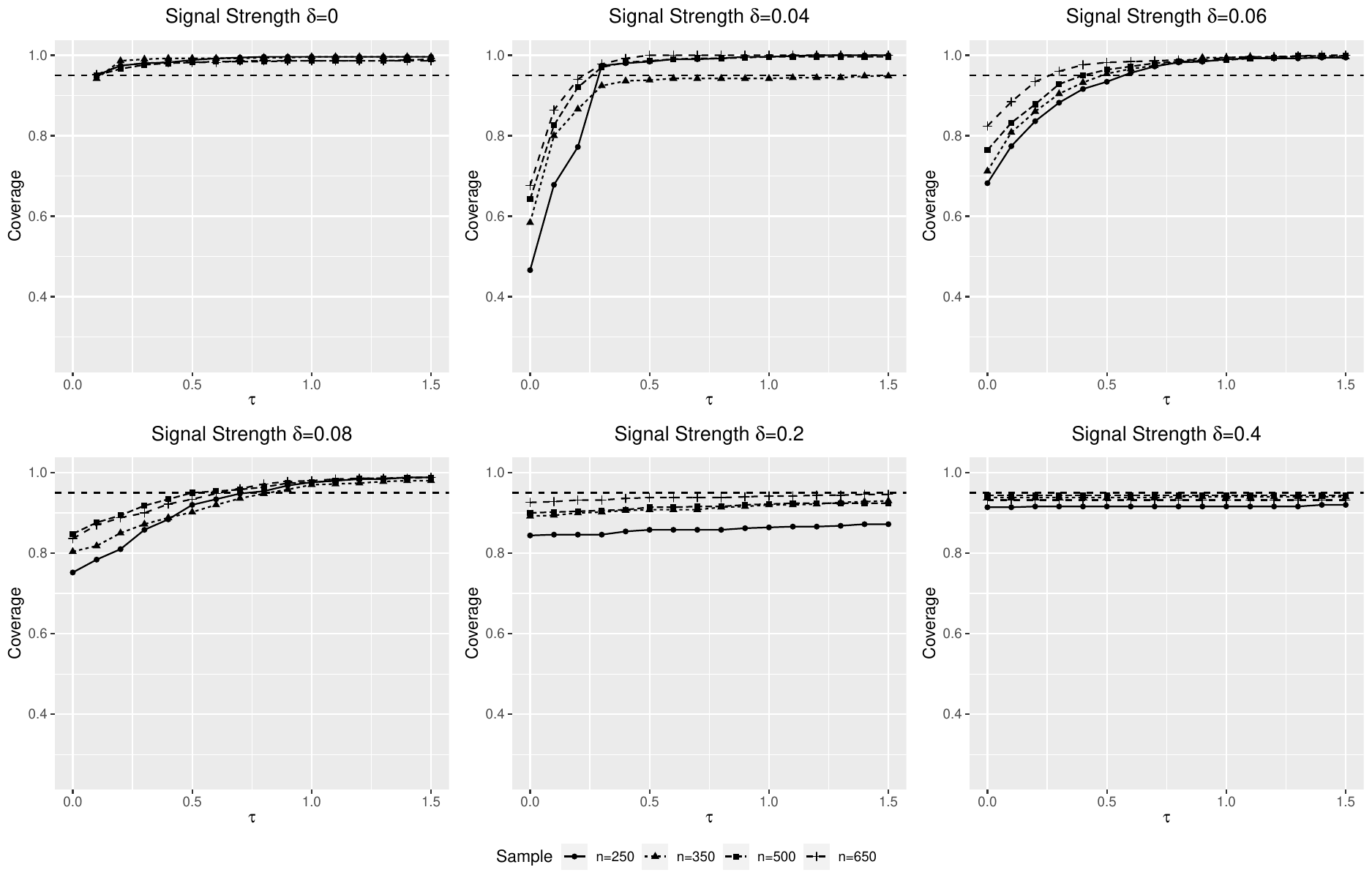}}%
    \caption{Dense alternative setting with a sparser signal: the dependence of Empirical Coverage on $\tau$.}%
    \label{fig: dense-cov-sparser}%
\end{figure}

We report the empirical rejection rate for $\phi_{\rm I}(\tau)$ and $\phi_{\Sigma}(\tau)$ in Figure \ref{fig: dense-det-sparser} and the coverage properties of the constructed confidence intervals ${\rm CI}_{\rm I}(\tau)$ and ${\rm CI}_{\Sigma}(\tau)$ in Figure \ref{fig: dense-cov-sparser}. By comparing Figure \ref{fig: dense-cov} and Figure \ref{fig: dense-cov-sparser}, we observe that for $\delta=0.4$, ${\rm CI}_{\rm I}(\tau)$ achieves the desired coverage level when the sample size $n$ reaches $350.$ This comparison shows that the inference problem is easier for a smaller sparsity level.

\subsection{Sample splitting}
\begin{figure}%
    \centering
    \subfigure[ERR of $\phi_{\rm I}(\tau)$ defined in \eqref{eq: test general}]{\includegraphics[width = 1\textwidth]{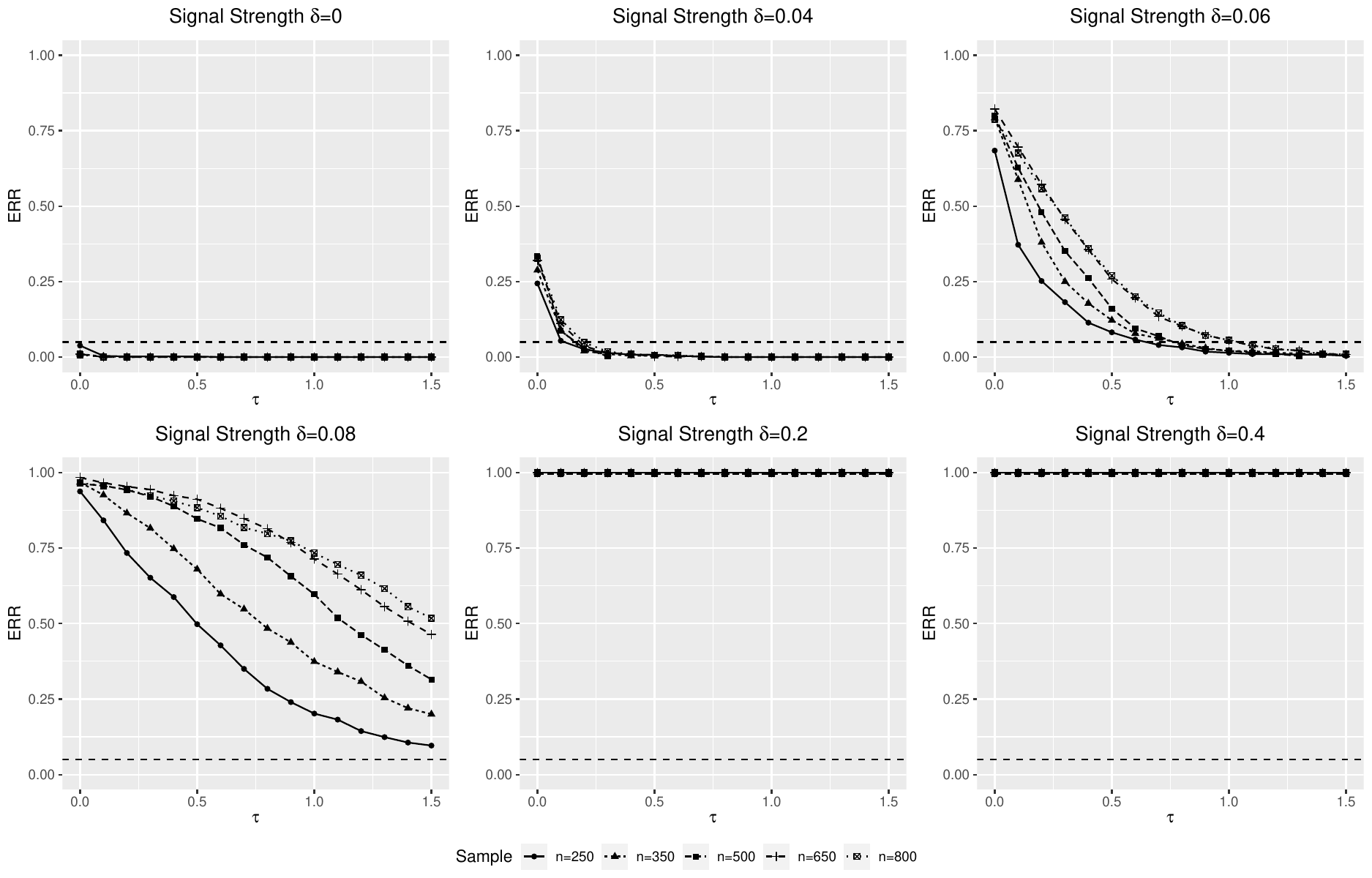}}%
    \qquad
	\subfigure[ERR of $\phi_{\Sigma}(\tau)$ defined in \eqref{eq: test Cov}]{\includegraphics[width = 1\textwidth]{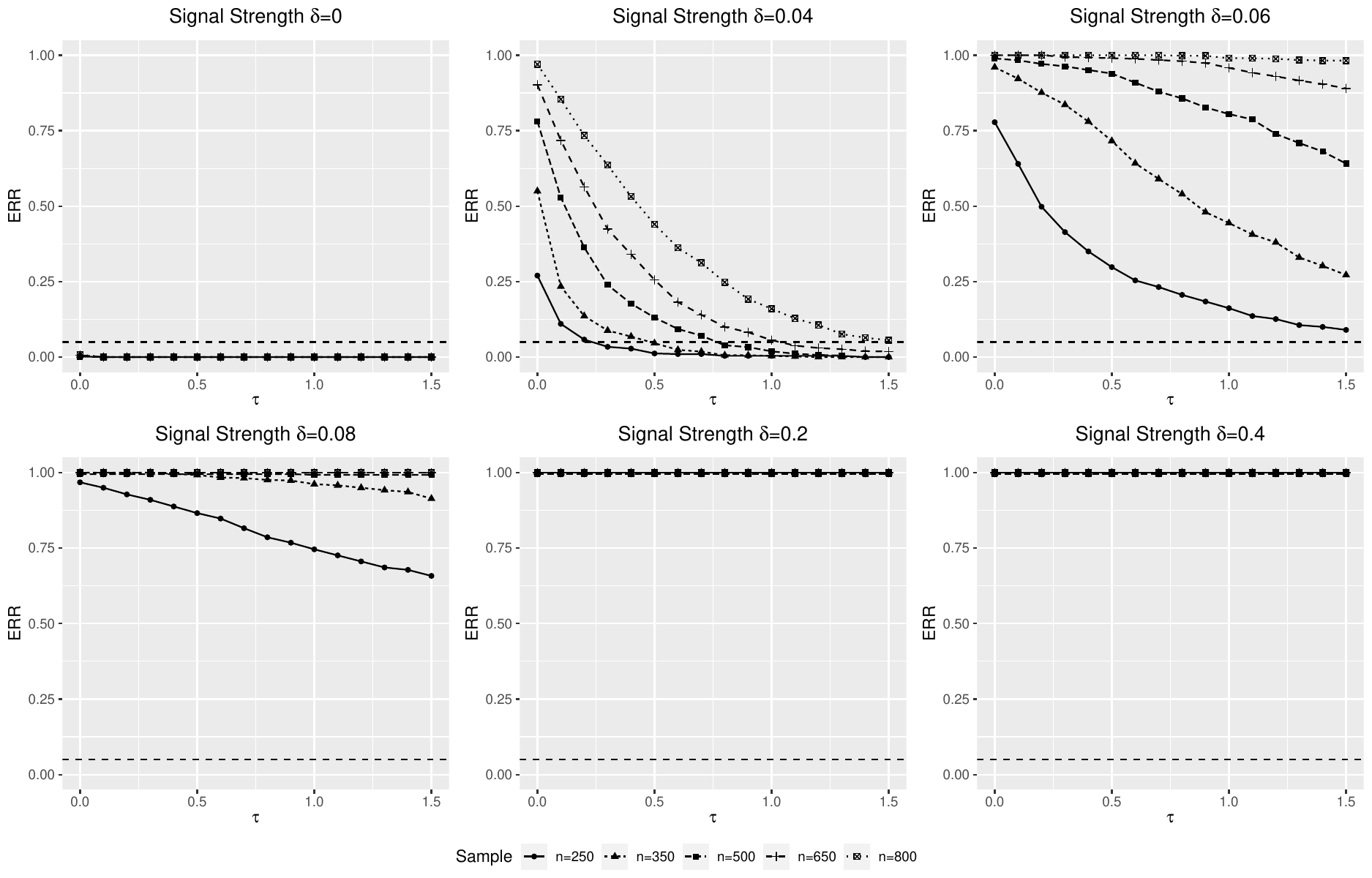}}%
    \caption{Sample splitting estimators for the dense alternative setting: the dependence of Empirical Rejection Rate (ERR)  on $\tau$.}%
    \label{fig: dense-det-split}%
\end{figure}

\begin{figure}%
    \centering
    \subfigure[Empirical Coverage of ${\rm CI}_{\rm I}(\tau)$ defined in \eqref{eq: CI general}]{\includegraphics[width = 1\textwidth]{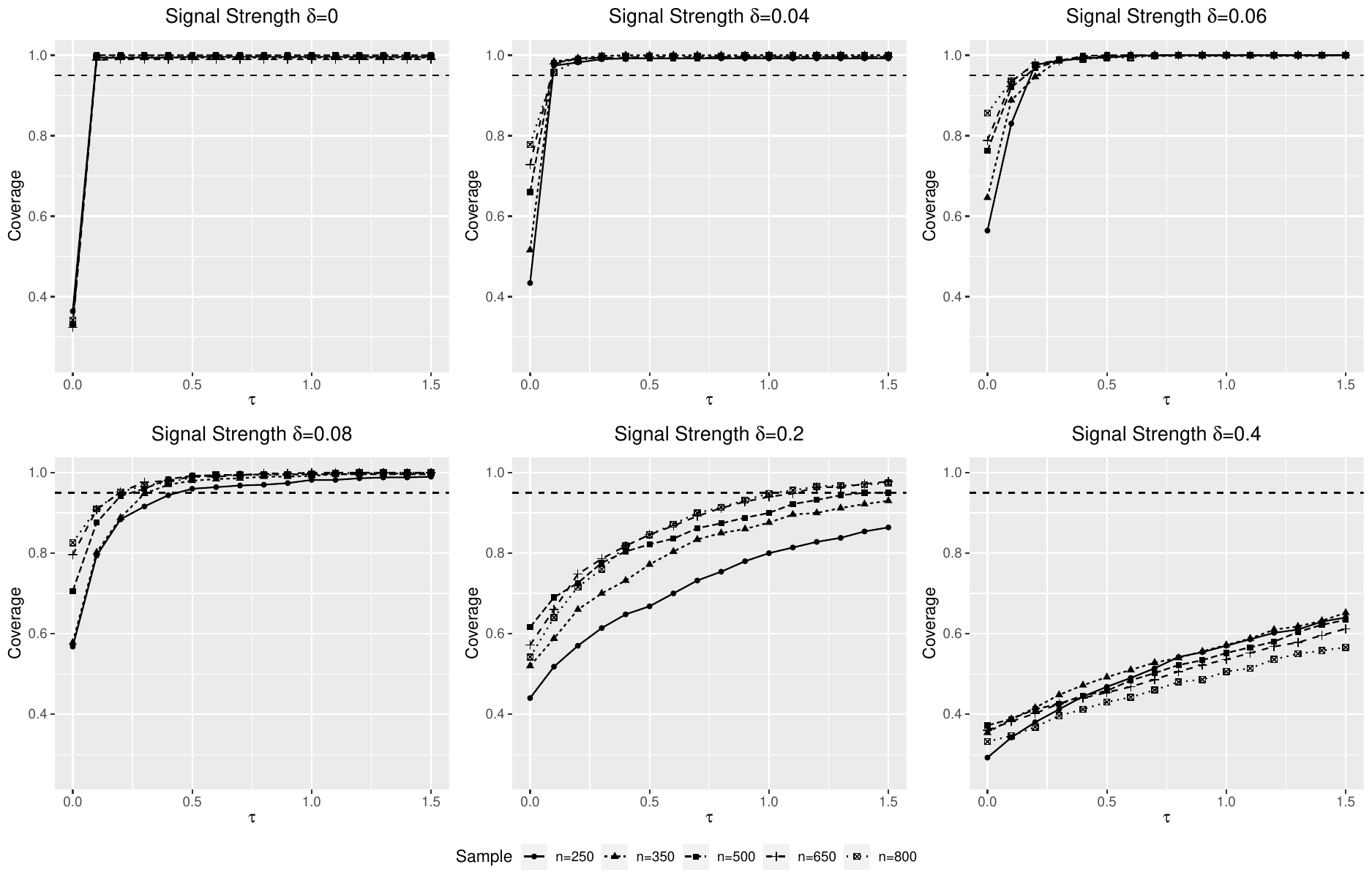}}%
    \qquad
	\subfigure[Empirical Coverage of ${\rm CI}_{\Sigma}(\tau)$ defined in \eqref{eq: CI Cov}.]{\includegraphics[width = 1\textwidth]{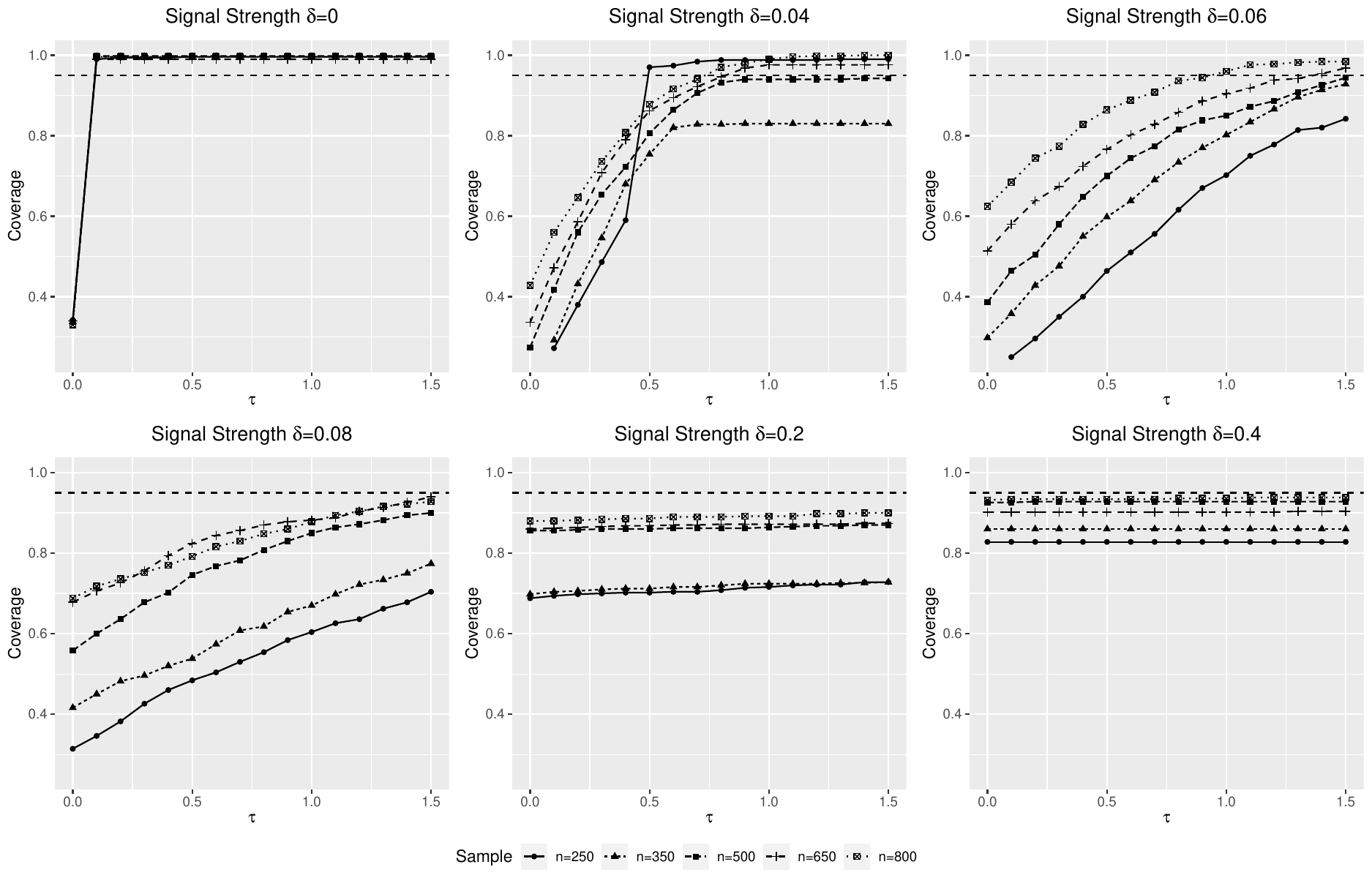}}%
    \caption{Sample splitting estimators for the dense alternative setting: the dependence of Empirical Coverage on $\tau$.}%
    \label{fig: dense-cov-split}%
\end{figure}

We implement the sample-splitting estimator detailed in \eqref{eq: projection Cov split}. Recall that this sample splitting estimator is mainly created to facilitate the proof. We take the same simulation settings as in Section \ref{sec: sim dense}. We vary the signal strength parameter $\delta$ over $\{0, 0.04,0.06, 0.08, 0.2,0.4\}$ and the sample size $n$ over $\{250,350,500,650,800\}$. We have examined the finite sample performance of the proposed method over different values of $\tau \in \{0,0.1,0.2,\cdots,1.4,1.5\}.$ We report the empirical rejection rate for $\phi_{\rm I}(\tau)$ and $\phi_{\Sigma}(\tau)$ in Figure \ref{fig: dense-det-split} and the coverage properties of the constructed confidence intervals ${\rm CI}_{\rm I}(\tau)$ and ${\rm CI}_{\Sigma}(\tau)$ in Figure \ref{fig: dense-cov-split}. In comparison to the results in Section \ref{sec: add dense}, we observe that the proposed tests are less powerful than the corresponding tests using full sample and the empirical coverage of the constructed confidence intervals is lower than that using the full data. This loss of efficiency is as expected as only half of the sample are used to construct the initial estimator and the other half are used to correct the bias. 

\subsection{High-dimensional setting with $p=4,000$}
We consider the group significance inference for a higher dimension with $p=4,000$. We vary the sample size across $\{500,1000,2000\}$ to mimic the dimension and the sample size of the colony growth data in Section \ref{sec: colony}. Regarding the generating parameters, we mimic Section \ref{sec: sim dense}, where the regression vector $\beta\in \R^{4000}$ is generated as $\beta_j=\delta$ for $25\leq j\leq 50$  and $\beta_j=0$ otherwise and generate the covariance matrix $\Sigma_{ij}=0.6^{|i-j|}$ for $1\leq i,j\leq 4,000$.  We consider the group significance test,
$
H_{0,G}: \beta_{i}=0 \; \text{for} \; i\in G, $ with  $G=\{30,31,\cdots,200\}$ or $G=\{30,31,\cdots, 60\}.$ We vary the signal strength parameter $\delta$ over $\{0, 0.08, 0.2,0.4\}$. 

\begin{figure}%
    \centering
    \subfigure[Test of $\beta_{G}=0$ with $G=\{30,31,\cdots,200\}$]{\includegraphics[scale=0.45]{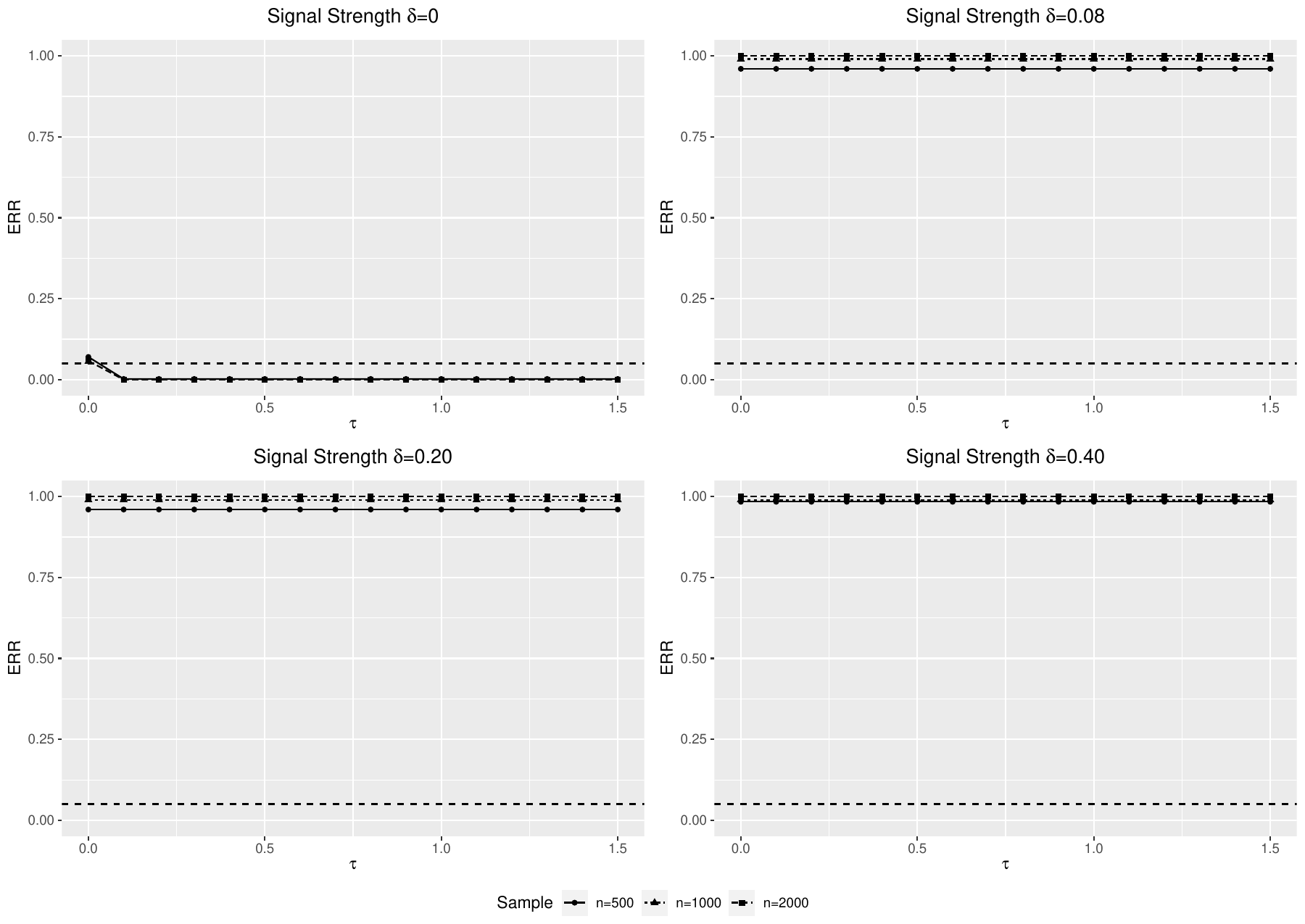}}%
    \qquad
	\subfigure[Test of $\beta_{G}=0$ with $G=\{30,31,\cdots,60\}$]{\includegraphics[scale=0.45]{DA-Largep-Sparse-Det}}%
    \caption{High-dimensional setting with $p=4,000$: the dependence of Empirical Rejection Rate (ERR) of $\phi_{\Sigma}(\tau)$ defined in \eqref{eq: test Cov} on $\tau$.}%
    \label{fig: largep-det}%
\end{figure}

We report the empirical rejection rate of $\phi_{\Sigma}(\tau)$ defined in \eqref{eq: test Cov} in Figure \ref{fig: largep-det}. The results are similar to the results for dense alternative presented in Section \ref{sec: sim dense} and \ref{sec: add dense}: for the null setting with $\delta=0$, the testing procedure controls the type I error for $\tau\geq 0.1$; for the alternative settings, the proposed test is powerful when $\delta$ reaches $0.08.$ The observations are uniform for testing both a larger group with $G=\{30,31,\cdots,200\}$ or a smaller group with $G=\{30,31,\cdots, 60\}.$
 
\begin{figure}%
    \centering
    \subfigure[Empirical Coverage of ${\rm CI}_{\rm I}(\tau)$ for $\beta_{G}^{\intercal}\Sigma_{G,G}\beta_{G}$ with $G=\{30,31,\cdots,200\}$]{\includegraphics[scale=0.45]{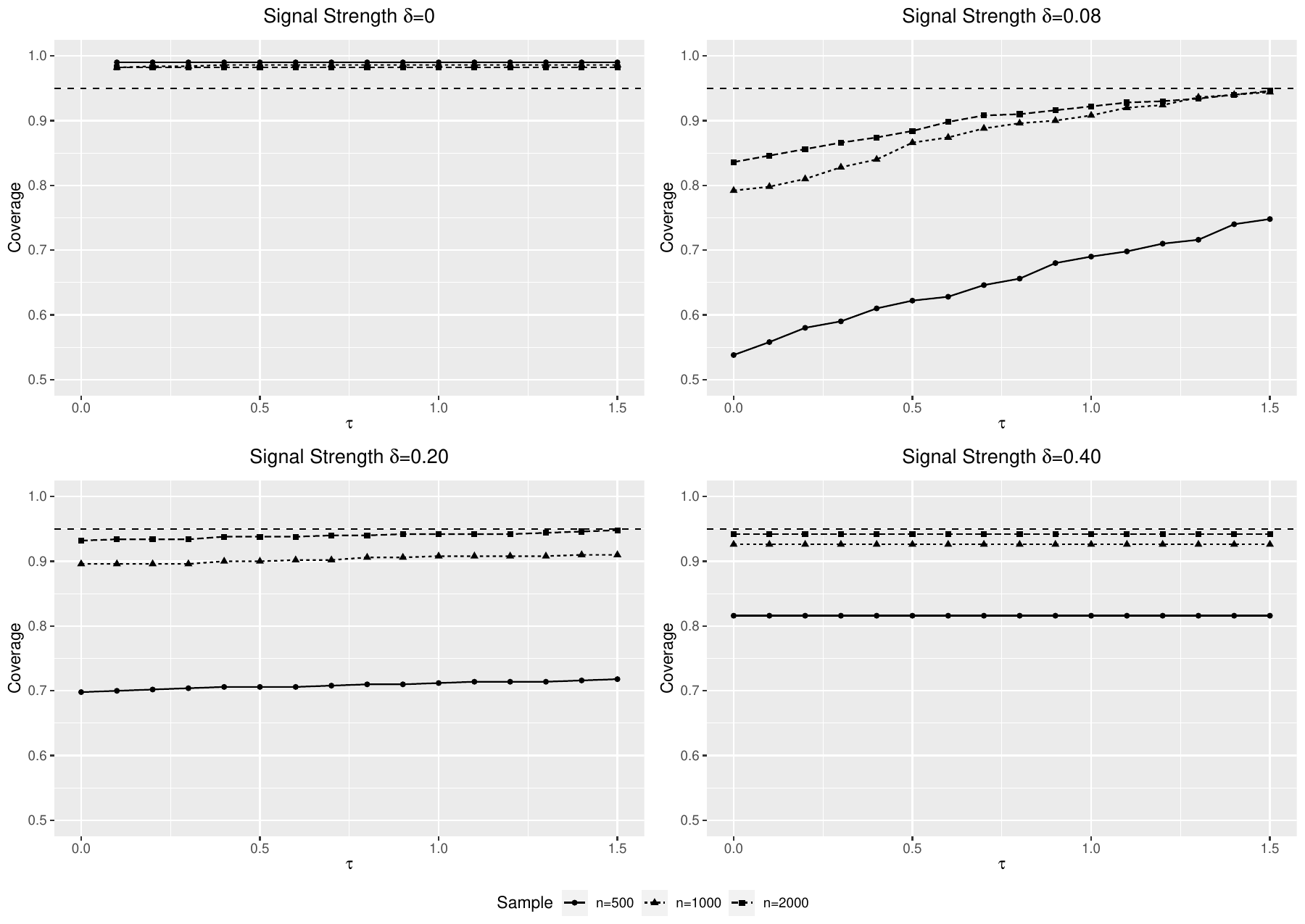}}%
    \qquad
	\subfigure[Empirical Coverage of ${\rm CI}_{\rm I}(\tau)$ for $\beta_{G}^{\intercal}\Sigma_{G,G}\beta_{G}$ with $G=\{30,31,\cdots,60\}$.]{\includegraphics[scale=0.45]{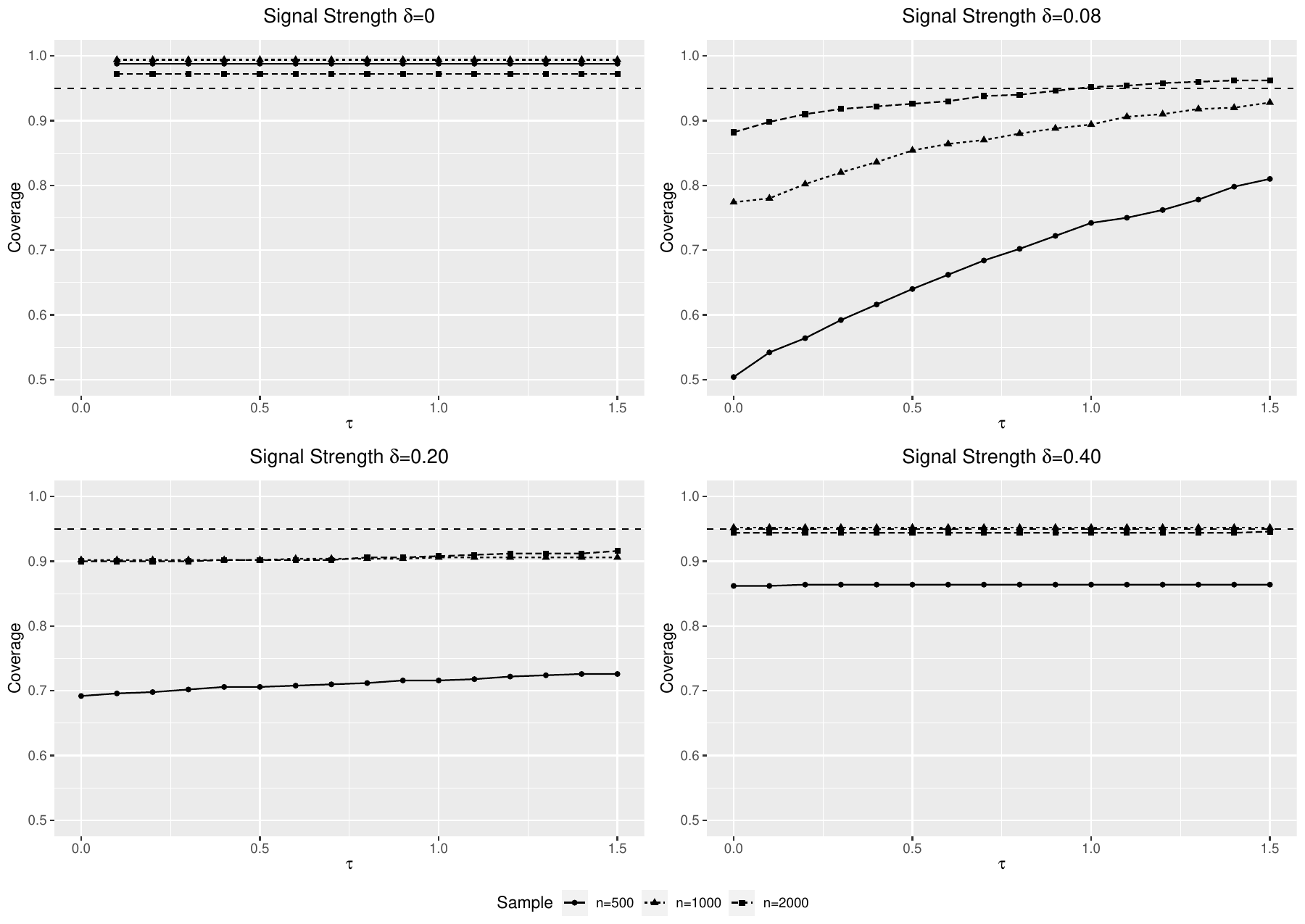}}%
    \caption{High-dimensional setting with $p=4,000$: the dependence of Empirical Coverage of ${\rm CI}_{\Sigma}(\tau)$ defined in \eqref{eq: CI Cov} on $\tau$.} 
    \label{fig: largep-cov}%
\end{figure}

We report the empirical coverage of the confidence interval construction ${\rm CI}_{\rm I}(\tau)$ defined in \eqref{eq: CI Cov} in Figure \ref{fig: largep-cov}. We can take $\tau=0.5$ or $1$ and in most settings, the empirical coverages are reliable when $n$ reaches $1,000$, which corresponds to the sample size for the colony growth data in Section \ref{sec: colony}. We note that, for $n=500$, the empirical coverages do not reach $95\%$ though the corresponding tests can still detect the signals when $\delta$ reaches $0.08.$

\subsection{Additional Real Data Analysis for the Yeast Colony Growth Data}

\begin{figure}%
    \centering
    \subfigure[Traits 1 until 23.]{\includegraphics[width = 1\textwidth]{plots-colony-bw-new-I}}%
    \qquad
	\subfigure[Traits 24 until 46.]{\includegraphics[width = 1\textwidth]{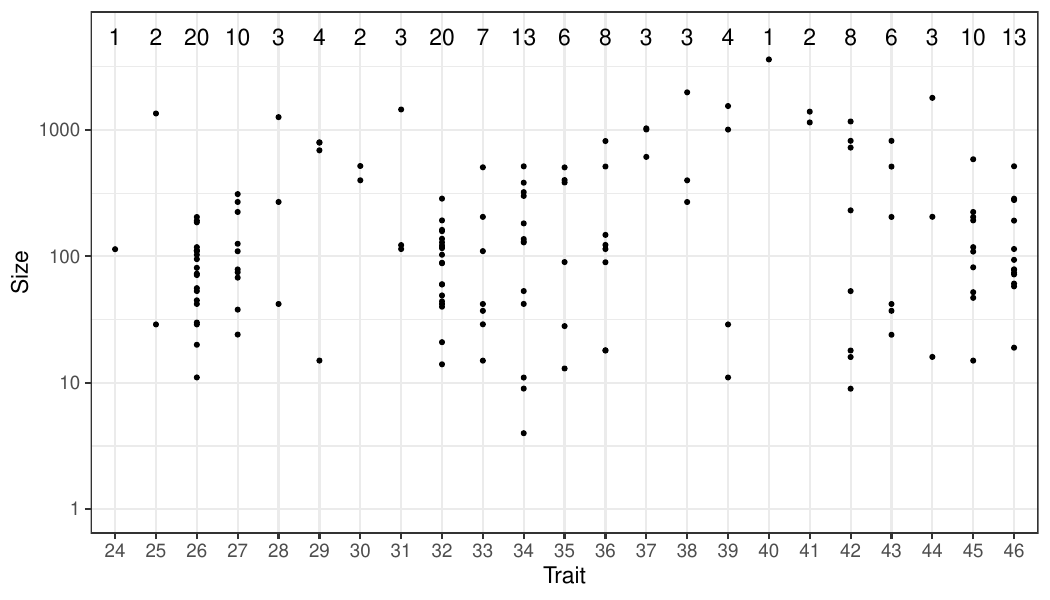}}%
    \caption{Size of significant groups (FWER level $5\%$) by applying
      the hierarchical procedure to each of the 46 traits of the Yeast
      Colony Growth data set. The number of significant groups is displayed
      on the top. The top panel is the same as Figure \ref{fig.colony} in the main paper.
      }%
    \label{fig.colony supp}%
\end{figure}

Figure \ref{fig.colony supp} describes the results for the colony data set from Section \ref{sec: colony}, but now for all 46 traits.

\subsection{Additional Real Data Analysis for the Riboflavin Data}
\label{sec: real data}

We apply the hierarchical procedure
on a data set about Riboflavin production with Bacillus Subtilis, 
made publicly available by \cite{bumeka13}. It consists of $n = 71$ samples of 
strains of Bacillus Subtilis with response being riboflavin (vitamin 
$\mbox{B}_2$) production rate and 
covariates measuring the log-expression levels of $p = 4088$ genes.

The hierarchical procedure using our group test $\phi_{\Sigma}(\tau=1)$ 
goes top-down through a hierarchical cluster
tree which was estimated using $1 - (\text{empirical correlation})^2$ as
dissimilarity measure and average linkage.
The results of the hierarchical procedure are displayed in Table
\ref{tab.ribo}. Hierarchical testing finds two single covariates, 
five small groups, and two large groups. 
The debiased estimator as implemented in the R package \texttt{\texttt{hdi}} \citep{hdipackage}
cannot reject any of the single covariates (when testing for all single covariates and adjusting p-values for controlling the FWER). Hence, the maximum test cannot reject the
global null hypothesis, implying that no significant group is found using hierarchical testing.

\begin{table}[ht]
\centering
\begin{tabular}{|l|l|}
  \hline
$p$-value & significant cluster \\ 
  \hline
6.113e-09 & LICT\texttt{\char`_}at, GLYQ\texttt{\char`_}at, PROA\texttt{\char`_}at, HEME\texttt{\char`_}at, PCP\texttt{\char`_}at, ... [5]    \\   
1.438e-05 & MURE\texttt{\char`_}at, YCGB\texttt{\char`_}at, YQEU\texttt{\char`_}at, SPOVC\texttt{\char`_}at, THDF\texttt{\char`_}at, ... [106]  \\ 
0.0002475 & YWAE\texttt{\char`_}at, YQZH\texttt{\char`_}at, YSGA\texttt{\char`_}at, YVDJ\texttt{\char`_}at, YQJU\texttt{\char`_}at, ... [2]     \\ 
$<$ 2.2e-16 & LYSC\texttt{\char`_}at, YDBH\texttt{\char`_}at, YDJL\texttt{\char`_}at, YDJK\texttt{\char`_}at, YHXA\texttt{\char`_}at, ... [2]     \\ 
0.0047295 &  YEBC\texttt{\char`_}at  \\                                                
1.768e-10 & CSPD\texttt{\char`_}at, OPUAB\texttt{\char`_}at, OPUAC\texttt{\char`_}at, OPUAA\texttt{\char`_}at, YLNA\texttt{\char`_}at, ... [924] \\ 
0.0001776 &  YOAB\texttt{\char`_}at       \\                                             
$<$ 2.2e-16 & XLYA\texttt{\char`_}at, YBFG\texttt{\char`_}at, XHLA\texttt{\char`_}at, XHLB\texttt{\char`_}at, XTMA\texttt{\char`_}at, ... [9]     \\ 
1.455e-11 & YXLE\texttt{\char`_}at, YXLF\texttt{\char`_}at, YXLC\texttt{\char`_}at, YXLD\texttt{\char`_}at, YXLG\texttt{\char`_}at    \\     
   \hline
\end{tabular}
\caption{Results from applying the hierarchical procedure to the Riboflavin 
data set (FWER level $5\%$). The number in square brackets indicates 
the number of covariates which are not displayed in the table, i.e.\ $[2]$ 
means that two covariates are not displayed for this group. 
} \label{tab.ribo}
\end{table}

\end{appendices}

\end{document}